\documentclass[preprint,showpacs,preprintnumbers,amsmath,amssymb] {revtex4}
\usepackage{graphicx}% Include figure files
\usepackage{dcolumn}% Align table columns on decimal point
\usepackage{bm}% bold math
\newcommand{\bee}{\begin{eqnarray}}
\newcommand{\eend}{\end{eqnarray}}
\newcommand{\rmd}{{\rm d}}
\newcommand{\rme}{{\rm e}}
\newcommand{\rmi}{{\rm i}}
\newcommand{\sch}{{Schr$\ddot{\rm o}$dinger equation\;}}

\begin{document}

%\preprint{APS/123-QED}

\title{Electric field of a pointlike charge in a strong magnetic field
 and ground state of a hydrogenlike atom}

\author{A.E. Shabad}

 \email{shabad@lpi.ru}
\affiliation{%
P.N. Lebedev Physics Institute, Moscow, Russia\\}

\author{V.V. Usov}
 \email{fnusov@wicc.weizmann.ac.il}
\affiliation{Center for Astrophysics, Weizmann Institute of Science,
Rehovot 76100, Israel\\}


\begin{abstract}
In an external constant magnetic field, so strong that the electron
Larmour length is much shorter than its Compton length, we consider
the modification of the Coulomb potential of a point charge owing to
the vacuum polarization. We establish a short-range component of the
static interaction in the Larmour scale, expressed as a Yukawa-like
law, and reveal the corresponding "photon mass" parameter. The
electrostatic force regains its long-range character in the Compton
scale: the tail of the  potential follows an anisotropic Coulomb
law, decreasing away from the charge slower along the magnetic field
and faster across. In the infinite-magnetic-field limit the
potential is confined to an infinitely thin string passing though
the charge parallel to the external field. This is the first
evidence for dimensional reduction in the photon sector of quantum
electrodynamics. The one-dimensional form of the potential on the
string is derived that includes a $\delta$-function centered in the
charge. The nonrelativistic ground-state energy of a hydrogenlike
atom is found with its use and shown not to be infinite in the
infinite-field limit, contrary to what was commonly accepted before,
when the vacuum polarization had been ignored. These results may be
useful for studying properties of matter at the surface of extremely
magnetized neutron stars.
\end{abstract}

\pacs{11.10.St, 11.10.Jj, 12.20.-m, 03.65.Pm}% PACS, the Physics and
%Astronomy
\maketitle

\section{Introduction}

 The fact that the vacuum, in which an external magnetic field is
present, is an optically  anisotropic medium, has been known,
perhaps, since the time when the nonlinearity of quantum
electrodynamics was first recognized: in a nonlinear theory
electromagnetic fields do interact with one another, provided that
the strength of at least one of them is of the order of or larger
than the characteristic value $B_0=m^2/e\simeq 4.4\times 10^{13}$G,
where $m$ and $e$ are electron mass and charge, respectively.
[Henceforth, we set $\hbar=c=1$ and refer to the Heaviside-Lorentz
system of units.] If the external field is strong, other fields
interact with it, the result of the interaction depending upon the
direction specified by the external field, hence the anisotropy.
Depending on the wave amplitude, the electromagnetic waves
propagation in this medium may be considered as a nonlinear process
\cite{bialyniska}, including the transformation of one photon into
two \cite{adler} or more photons, or taken in the linear
approximation with respect to the amplitude. In the latter case, the
second-rank polarization tensor is responsible for the properties of
the medium. In the kinematic domain where the photon absorption
processes like electron-positron pair creation are not allowed, the
polarization tensor is symmetric and real, and the medium is
transparent and birefringent \cite{footnote}. In the absorption
domain the medium is dichroic \cite{heyl}. The limit of low
frequency and momentum belongs to the transparency domain and
corresponds to a constant anisotropic dielectric permeability of the
medium. In this limit the polarization tensor may be obtained by
differentiations with respect to the fields of an effective
Lagrangian, calculated on the class of constant external electric
and magnetic fields. For small values of these fields \cite{blp} and
for
 extremely large  \cite{heyl} fields the polarization operator
was in this way considered using the  effective Lagrangian of
Heisenberg-Euler calculated \cite{euler} within the one-loop
approximation. (The two-loop calculations are also available
\cite{lebedev}.) The knowledge of this limit is useful for studying
the dielectric screening of the fields that are (almost) static and
(almost) constant in space. For more general purposes, however, this
limit is not sufficient, and one should calculate the polarization
tensor directly, using the Feynman diagram technique of the Furry
picture in the external magnetic field. On the photon mass shell,
i.e. when the photon energy, $k_0$, and 3-momentum, $\bf k$, are
related by the free vacuum dispersion law $k_0^2={\bf k}^2,$ such
calculations were done by Adler \cite{adler2} and Constantinescu
\cite{constantinescu}. The results obtained are appropriate for
handling the photon propagation in weakly dispersive medium, when
the dispersion law does not essentially deviate from its vacuum
shape. The polarization operator for the case of general relation
between the photon mass and momentum was calculated by Batalin and
Shabad \cite{batalin}, Tsai \cite{tsai}, Baier et {\em al.}
\cite{baier}, and Melrose and Stoneham \cite{melrose1}. This gave
the possibility of studying the photon propagation \cite{annphys}
under the conditions where the deviation from the vacuum dispersion
law may be very strong either due to the phenomenon of the cyclotron
resonance in the vacuum polarization \cite{nuovcimlet} - this
phenomenon is responsible for the effect of photon capture by a
magnetic field \cite{nature,ShUs,wunner} - or due to magnetic
fields, much larger than $B_0$
\cite{kratkie,shabtrudy,mikheev,japan}, or due to the both
circumstances (see Ref. \cite{zhetf} where the photon capture effect
was extended to low frequencies for extra-large fields).

Although much work has been devoted  to  study of electromagnetic
wave propagation in the magnetized vacuum, problems of electro- and
magneto-statics in this medium did not attract sufficient attention,
save Refs. \cite{loskutov1, skobelev}, where corrections to the
Coulomb law were found when these are small: for $B/B_0 \ll 1$ in
\cite{loskutov1}, or at large distances from the source for $1\ll
B/B_0\ll 3\pi\alpha^{-1}$ in \cite{skobelev}, where $\alpha =
e^2/4\pi=1/137$ in the fine-structure constant. Here we proceed with
an investigation of some electrostatics in the presence of a strong
external magnetic field in the vacuum to find that for sufficiently
large $b\equiv B/B_0 \gg 1$ the electric field produced by a
pointlike charge at rest may be significantly modified by the vacuum
polarization, the modification being determined by the
characteristic factor $\alpha b$. We note first of all that
expressions for the dielectric permeability of the magnetized vacuum
obtained from the Heisenberg-Euler Lagrangian are applicable only as
far as the fields slowly varying in the space are concerned.
Otherwise, the spatial dispersion becomes important. For this
reason, when considering the electric field produced by a pointlike
electric charge in the present paper, we address again to the
polarization tensor calculated off-shell in
\cite{batalin,tsai,baier,melrose1,annphys}, taking it in the static
limit $k_0=0$, but keeping the dependence on $\bf k$. The
corresponding spatial dispersion effects will be essential for
getting some important features of the modified Coulomb potential.

Using the tensor decomposition of the polarization operator and
the photon Green function, established in \cite{batalin,annphys}
in an approximation-independent way, we find that photons of only
one polarization mode (mode-2 in the nomenclature of these
references, see below) may be carriers of electrostatic force.
This is in agreement with the fact that the electromagnetic field
of these photons is, in the static limit, purely electric and
longitudinal. The photons of the other two modes mediate in this
limit the magneto-static field of constant currents.

In magnetic fields $B\gg B_0$, which we are dealing with when
describing the static field, produced in the magnetized vacuum by a
point electric charge, the electron Larmour length $L_{\rm B}\equiv
(eB)^{-1/2}=\lambda_{\rm C}b^{-1/2}$ is much less than the electron
Compton length $\lambda_{\rm C}=m^{-1}$. Therefore, two different
scales occur in the problem: the Larmour scale and the Compton
scale.

A simplifying expression for the mode-2 eigenvalue of the
polarization operator is used, valid for such fields. It was first
obtained by Loskutov and Skobelev \cite{skobelev} within a special
two-dimensional  technique intended for large fields, and by Shabad
\cite{kratkie}, and Melrose and Stoneham \cite{melrose1} as the
asymptote of the mode-2 eigenvalue calculated
\cite{batalin,tsai,baier,melrose1,annphys} in the one-loop
approximation (see \cite{zhetf} for the detailed derivation of the
large-field asymptotic behavior.) The most important, now widely
accepted, fact about this asymptotic behavior (see, {\em e.g.,} the
monographs \cite{shabtrudy,dittrich,kuznetsov}) is that the mode-2
eigenvalue contains a term linearly growing with the magnetic field,
seen already \cite{heyl} if one deals with nondispersive small
momentum approximation, inferrable from the Heisenberg-Euler
Lagrangian. It is sometimes expected that this term - it appears in
the denominator of the photon propagator and hence of the expression
for the potential - should lead to suppression of the interaction
mediated by mode-2 photons. In a different problem we already had an
opportunity to show that this is not exactly the case \cite{prl}. In
\cite{zhetf} the impact of the linear term on the refractive index
was considered. In the present paper we demonstrate that this term
is also crucial for the most important features of the potential of
a point charge.

Correspondingly to the two scales inherent in the problem, the
 potential is divided into two additive parts, out of which the
 first one, called short-range, decreases exponentially at
 distances $r$ of a few Larmour lengths from the source, but retains
 the usual $q/4\pi r$ singularity near the origin $r=0$, where the charge $q$ is
 placed. (The anisotropy shows itself no sooner than in the third term of
 the Laurent expansion around the singular point $r=0$ - see
 Appendix I.)
  The second part, called long-range, slowly decreases away from
  the charge following an anisotropic Coulomb law, but remains finite close to the charge.
   The linear term mentioned above, is
  responsible for the fact that a scaling regime of the short-range part
  occurs, characterized by a comparatively simple universal function,
independent of the magnetic field. This function gives the
potential of a point charge in the energy units of $L^{-1}_{\rm
B}= (eB)^{1/2}$ as a dimensionless function of the space
coordinates in the units of the electron Larmour length $L_{\rm
B}$. Excluding the closest vicinity of the  charge, its form
coincides with the Yukawa law [see Eq. (\ref{yuk}) below]
characterized by the dimensionless mass parameter $2\alpha/\pi$
(which is the topological mass of the two-dimensional Schwinger
electrodynamics \cite{schwinger},
%"effective photon mass" %discussed in \cite{kuznets},
measured in inverse Larmour lengths).
 Thus, this mass governs the  exponential (isotropic)
 decrease of the potential away from its source at the distances,
  large in the Larmour scale.
 In the
 formal limit of infinite magnetic field the short-range part
 becomes the $\delta$-function with its center in the charge.
 As one moves
 farther from the source, the Yukawa decrease ceases, and the
 potential  coincides with its long-range part. It approaches, for
 distances from the charge, large in the Compton scale,
  the anisotropic Coulomb shape of the
 form of Eq. (\ref{llargex}) (that might have been also derived
 disregarding the spatial dispersion). The law of
 decreasing along the magnetic field is unaffected by the latter,
 the decrease is the fastest in the direction orthogonal to the
 magnetic field. Thus, the linearly growing term in the mode-2
 eigenvalue of the polarization operator leads, first, to the faster
 decrease of the potential in the direction across the magnetic
 field for large distances, and, second, to its steeper shape for small
 distances. This may be recognized as suppression, indeed. On the other
 hand, the long-distance behavior along the magnetic field, as
 well as the standard Coulomb singularity near the source \cite{to avoid} do not
sense the magnetic field at all, no matter how strong it is.

Perhaps, the most interesting feature of the potential produced
 by a point charge is that, as the external magnetic field tends to
 infinity, the whole potential becomes concentrated inside an
 infinitely thin string that includes the charge and is directed
 parallel to the magnetic field. The electric lines of force produced
 by the charge are gathered
 inside the string. The string is the $b=\infty$ limit of an ellipsoidal
 equipotential surface. The potential
 along the string as a function of the longitudinal distance from the charge
 is just the infinite-magnetic-field limit of the
 long-range part of the potential (see Fig. \ref{fig:6} in Section IV)
 plus the $\delta$-function
 contribution from the short-range part. The string potential
  first  grows with distance logarithmically and linearly
 (starting with negative values) in-between the Larmour and Compton distances
 and hence provides "confinement" in this scale. For the string formation, again,
 the above-discussed term, linearly growing with the magnetic field, is responsible.
 To conclude about its presence, a consideration of the Heisenberg-Euler Lagrangian
might have been sufficient.
 However, for calculating the string potential, the effect of spatial dispersion was important.
 In contrast to the
 quark-antiquark string in the lattice QCD, the string potential of the present paper
 stops growing  after reaching the Compton distances from the charge and approaches
 zero following the Coulomb law $1/4\pi r$ in accord with the fact that the infrared
 singularity in QED is milder than in QCD and insufficient to provide the infra-red
 custody.

 The appearance of the string notifies the reduction of QED to two
 dimensions (one time-one space) in the photon sector in the
 infinite-magnetic-field limit, which implies a new result \cite{ritus},
 because what was known before was the reduction to two dimensions in the
 electron sector. The latter circumstance is a common knowledge and is well
 understood referring to the fact that electrons are confined to
 the lowest Landau level, so only one degree of freedom - that
 along the magnetic field - survives to remain dynamical. For
 instance, it was demonstrated in \cite{prl}, that
  the Bethe-Salpeter equation
 describing the interaction between electrons and positrons
 acquires in this limit fully Lorentz-covariant form in the two-dimensional
 space with the metrics (1,-1).

 Analogously, the nonrelativistic \sch for an electron in the field
 of the nucleus of a hydrogenlike atom is known to become a
 differential equation with respect to the longitudinal distance
 between the two particles. Unless the vacuum polarization is
 taken into account, the standard result \cite{elliott} is that
  due to the singularity
 of the Coulomb potential in the origin, the ground-state
 energy in this problem tends to negative infinity as the magnetic
 field unlimitedly grows. The conclusion of the present paper is that if
 the string potential obtained is used in the \sch the ground
 state energy remains finite just because the
  singularity of
 the string potential in the origin has the $\delta$-function
 character.

Another conclusion concerns the critical nucleus charge $Z_{\rm cr}$
making the threshold of its instability manifested in spontaneous free
 positron production. The known fact here \cite{semikoz} is that $Z_{\rm cr}$
becomes reasonably smaller than $\alpha^{-1}=137$ for large magnetic fields.
This result depends on the same unboundedness
from below of the energy spectrum of the Dirac operator caused by the same
Coulomb singularity of the static potential. Therefore, it should be revised
if the vacuum polarization of the Coulomn potential is taken into account.

The paper is organized as follows. In Section II we present an
approximation-independent form of the potential of a point charge
in terms of the mode-2 component of the photon propagator and
define the division of the potential corresponding to the one-loop
polarization operator in the asymptotical region of large magnetic
fields into the short- and long-range parts.  In Section III we
consider the
short-range part as determined by an %Three terms of its asymptotic
expansion near the
%source are explicitly written.
 universal function corresponding to  the scaling  regime. It is obtained
 as a one-fold
integral. We also establish the $\delta$-function character of the
short-range part in the limit of infinite magnetic field. In Section
IV the anisotropic Coulomb law is obtained for large distances for
the long-range part of the potential  by applying
 mathematical means, differnt for different remote
regions in the space. The limiting, $b=\infty$, form of the
long-range part is studied for short and long distances on the
string. In Section V we estimate the limiting value of the ground
state energy of a hydrogenlike atom by considering the \sch with the
vacuum-polarization-modified potential and using the
shallow-well-potential method. Also a perturbation correction to the
ground state valid for the fields in the range $1\ll
b\ll2\pi/\alpha$ is found. A role the radiative modification of
Coulomb potential may play when the Dirac equation is used is
discussed. In Section~VI we briefly discuss possible applications of
our results to physics of strongly magnetized neutron stars. In
Appendix I serving Section II we derive the asymptotic expansion of
the potential near the point $r=0$ and study the coefficients in
this expansion as functions of the magnetic field. Also an analog of
Uehling-Silber \cite{blp} correction to the Coulomb potential valid
in the interval $1\ll b\ll2\pi/\alpha$ is derived. In Appendix II we
deal with a simplified potential that models the short-range
Yukawa-like potential and also has the $\delta$-function as a
limiting form. In this case explicit solution of the \sch can be
obtained with the use of the method of Refs. \cite{haines}. The
finiteness of the ground energy is demonstrated.

Some results of the present paper were shortly reported  in our
previous paper \cite{PRL07}. See also the preliminary publication
\cite {arxive}, more detailed in certain points.

\section{General representation for the static potential of a pointlike  charge:
One-loop approximation in the large magnetic field domain}
Electromagnetic 4-vector potential produced by the 4-current
$j_\nu(y)$ is \bee\label{4-pot} A_\mu(x)=\int
D_{\mu\nu}(x-y)j^\nu(y)\rmd^4y,\quad \mu,\nu=0,1,2,3.\eend Here
$x$ and $y$ are 4-coordinates, and $D_{\mu\nu}(x-y)$ is the photon
Green function in a magnetic field in the coordinate
representation. The metric in the Minkowski space is defined so
that diag $g_{\mu \nu}=(1,-1,-1,-1)$. Eq. (\ref{4-pot}) defines
the 4-vector potential with the arbitrariness of a free-field
solution, including the gauge arbitrariness. If the photon Green
function is chosen as causal, only the gauge arbitrariness
remains.

The current of a pointlike static charge $q$, placed in the point
${\bf y}=0$ is\bee\label{current} j^\nu(y)=q\delta_{\nu
0}\delta^3({\bf y}).\eend Hence\bee\label{4-pot2} A_\mu({\bf
x})=q\int_{-\infty}^\infty D_{\mu 0}(x_0-y_0,{\bf x})\rmd y_0\nonumber\\
=q\int_{-\infty}^\infty  D_{\mu 0}(x_0+y_0,{\bf x})\rmd
y_0=q\int_{-\infty}^\infty D_{\mu 0}(y_0,{\bf x})\rmd y_0.\eend
This 4-vector potential is also static.

If there is no magnetic field, and the photon propagator is free
and taken in the Feynman gauge (with the pole handled in a causal
way) \bee\label{free}D_{\mu\nu}(x-y)=D_{\mu\nu}^{\rm C}
(x-y)\equiv\frac {g_{\mu\nu}}{\rmi 4\pi^2(x-y)^2},\eend  only the
zeroth component of the 4-vector potential is
present:\bee\label{free2} A_0^{\rm C}({\bf
x})=q\int_{-\infty}^\infty D^{\rm C}_{\mu 0}(y_0,{\bf x})\rmd
y_0=\frac{q}{\rmi 4\pi^2}\int_{-\infty}^\infty \frac{\rmd
y_0}{y_0^2-{\bf x}^2}=\frac 1{4\pi}\frac{q}{|{\bf x}|}.\eend This
is the Coulomb potential in the Heaviside-Lorentz system of units
used throughout.

Let there be an external magnetic field $B$ directed along axis 3
in the Lorentz frame where the charge $q$ is at rest in the origin
${\bf x}=0$ , and no external electric field exists in this frame.
Call this frame special. Define the Fourier transform as
\begin{eqnarray}\label{fourier} D_{\mu\nu}(x)=\frac
1{(2\pi)^4}\int \exp ({\rm i}kx) D_{\mu\nu}(k)~{\rm d}^4k,\quad
\mu,\nu=0,1,2,3.
\end{eqnarray} Then (\ref{4-pot2}) becomes\begin{eqnarray}\label{potmag}
%\hspace{-3cm}
A_\mu({\bf x})=\frac{q}{(2\pi)^4}\int \exp [\,{\rm i}(k_0y_0-{\bf
kx})] D_{\mu 0}(k)~{\rm d}^4k\,\rmd y_0 = \frac{q}{(2\pi)^3}\int
D_{\mu 0}(0,{\bf k})\exp(-\rmi{\bf kx})\,\rmd^3k.
\end{eqnarray}

The four 4-eigenvectors $\flat^{(a)}_\nu$, $a=1,2,3,4$, of the
polarization tensor $\Pi_{\mu\nu}$ \cite{batalin, annphys,
nuovcimlet} take in the special frame (and arbitrary
normalization) the form - the components are counted downwards as
$\nu= 0,1,2,3$ -
\begin{eqnarray}\label{b}
\flat_\nu^{(1)}=k^2\left(\begin{tabular}{c}0\\$k_1$\\$k_2$\\0\end{tabular}\right)_\nu+
k_\perp^2\left(\begin{tabular}{c}$k_0$\\$k_1$\\$k_2$\\$k_3$\end{tabular}\right)_\nu,\quad
\flat_\nu^{(2)}=\left(\begin{tabular}{c}$k_3$\\0\\0\\$k_0$\end{tabular}\right)_\nu,\nonumber\\
\flat_\nu^{(3)}=\left(\begin{tabular}{c}0\\$k_2$\\$-k_1$\\0\end{tabular}\right)_\nu,\quad
\flat_\nu^{(4)}=\left(\begin{tabular}{c}$k_0$\\$k_1$\\$k_2$\\$k_3$\end{tabular}\right)_\nu,
\end{eqnarray} Among them there is only one, whose zeroth component %(the upper raw in (\ref{\flat}))
survives the substitution $k_0=0$. It is $\flat_\nu^{(2)}$. This
implies that out of the four ingredients of the general
decomposition of the photon
propagator\begin{eqnarray}\label{decomposition} D_{\mu\nu}(k)=
\sum_{a=1}^4 \mathcal{D}_a(k)~\frac{\flat_\mu^{(a)}~
\flat_\nu^{(a)}}{(\flat^{(a)})^2},\nonumber\\
\mathcal{D}_a(k)=\left\{\begin{tabular}{cc}$-[k^2+\kappa_a(k)]^{-1},$
&$\quad a$=\{1,\,2,\,3\}\,,\\arbitrary,
&$a$=4\,,\end{tabular}\right.
\end{eqnarray}where $\kappa_a(k)$ are scalar eigenvalues of the
polarization tensor:
\begin{eqnarray}\label{eigen}\Pi_\mu^{~\nu}(k)~\flat^{(a)}_\nu=\kappa_a(k)~\flat^{(a)}_\mu,
\qquad \kappa_4(k)=0\,,\end{eqnarray} only the term with $a=2$,
$\mathcal{D}_2(k)\flat_{\mu}^{(2)}\flat_{\nu}^{(2)}/(\flat^{(2)})^2$,
participates in (\ref{potmag}), i.e. only  mode-2 (virtual)
photons may be a carrier of electro-static interaction, and not
photons of modes 1 and 2, nor the purely gauge mode 4. Bearing in
mind that $(\flat^{(2)})^2=k_3^2-k_0^2$, we have\bee\label{A0}
A_0({\bf x})=\frac{q}{(2\pi)^3}\int\frac{\rme^{-\rmi{\bf
kx}}\rmd^3k}{{\bf
k}^2-\kappa_2(0,k_3^2,k_\perp^2)},~~A_{1,2,3}({\bf x})=0\,.\eend
Here $k_\perp^2=k_1^2+k_2^2$. Thus, the static charge gives rise
to electric field only, as it might be expected. The gauge
arbitrariness in the choice of the photon propagator
$D_4(k)\flat_{\mu}^{(4)}\flat_{\nu}^{(4)}=\mathcal{D}_4(k)k_{\mu}k_{\nu}$
indicated in (\ref{decomposition}) has no effect in (\ref{A0}).
Certainly, the potential (\ref{A0}) is defined up to gauge
transformations.

The result that only mode-2 photons mediate electrostatic
interaction may be understood, if we inspect electric and magnetic
components of the fields of the eigen-modes obtained from their
4-vector potentials (\ref{b}) in the standard way: ${\bf
e}^{(a)}=k_0{\bf b}^{(a)}-{\bf k}\flat_0^{(a)}$, ${\bf
h}^{(a)}={\bf k}\times{\bf b}^{(a)}.$ These are \cite{annphys}
\bee\label{5} {\bf e}^{(1)} =-\frac{{\bf k_\perp}}{k_\perp}k_0,
\hspace{35mm} {\bf h}^{(1)} = (\frac{\bf k_\perp}{k_\perp} \times
{\bf k}_3), \eend\bee\label{5b} {\bf e}^{(2)}_\perp = {\bf
k}_\perp k_3, \hspace{7mm}{\bf e}^{(2)}_3 = \frac{{\bf k}_3}{k_3}
(k_3^2-k_0^2),\hspace{3mm} {\bf h}^{(2)} = -k_0\left({\bf
k_\perp}\times \frac{{\bf k_3}}{k_3}\right),\eend\bee\label{5c}
{\bf e}^{(3)} = - k_0 \left(\frac{{\bf
k_\perp}}{k_\perp}\times\frac{{\bf
k_3}}{k_3}\right),\hspace{3mm}{\bf h_\perp}^{(3)}=-\frac{\bf
k_\perp}{k_\perp} k_3,\hspace{3mm} {\bf h_3}^{(3)}= \frac{\bf
k_3}{k}_3 k_\perp,  \eend where the cross stands for the vector
product, and the boldfaced letters with subscripts $3$ and $\perp$
denote vectors along the directions, parallel and perpendicular to
the external magnetic field, respectively.

The photon energy and momenta here are not, generally, related by
any dispersion law. Therefore, we may discuss polarizations of
virtual, off-shell photons - carriers of the interaction - basing
on Eqs. (\ref{5})-(\ref{5c}). The electric field $\bf e$ in mode 1
is parallel to $\bf k_\perp$, in mode 2 it lies in the plane
containing the vectors $\bf k, B$, in mode 3 it is orthogonal to
this plane, $i.e.$ mode 3 is always transversely polarized, ${\bf
e}^{(3)}{\bf k}=0$.   For the special case of the virtual photon
propagation transverse to the external magnetic field, $k_3=0$,
(this reduces to the general case of propagation under any angle
$\theta\neq 0$ by a Lorentz boost along the external magnetic
field), mode 2 is transversely polarized , ${\bf e}^{(2)}{\bf
k}=0$,  as is always the case for mode 3. Mode 1 for transverse
propagation, $k_3=0$, is longitudinally polarized, ${\bf}
e^{(3)}\times {\bf k}=0$, and its magnetic field is zero. The
lowest-lying cyclotron resonance of the vacuum polarization
\cite{nuovcimlet}, the one that corresponds to the threshold
$k_0^2-k_3^2=4m^2$ of creation of the pair of electron and
positron in the lowest Landau state each, belongs to mode 2. It
gives rise to the photon capture effect with the photon turning
into a free \cite{nature} or bound \cite{ShUs, wunner}
electron-positron pair. Another consequence of the cyclotron
resonance is that a real photon of mode 2 undergoes the strongest
refraction in the large magnetic field limit \cite{zhetf} even if
its frequency is far beyond the pair production threshold.

In the static limit $k_0=0$ the magnetic field in mode 2
disappears, ${\bf h}^{(2)}=0$, while its electric field is
collinear with ${\bf k},$ ${\bf e}^{(2)}={\bf k}.$ It becomes a
purely longitudinal virtual photon. Unlike  mode 2, in modes 1 and
3 in the static limit $k_0=0$ only the magnetic fields survive:
${\bf e}^{(1,2)}=0$, ${\bf h}^{(1)}={\bf k}_\perp\times{\bf B},$
${\bf h}_\perp^{(3)}=-{\bf k}_\perp k_3,$ $h_3^{(3)}=k_\perp^2$,
${\bf h}^{(1,3)}{\bf k}=0$. (Here normalizations are arbitrary and
kept fixed only within the same mode). A virtual mode-1 photon
carries magneto-static interaction. It is responsible for magnetic
field produced by a current flowing through a straight-linear
conductor oriented along the external magnetic field. In
accordance with the above formula for ${\bf h}^{(1)}$ its magnetic
field is orthogonal both to ${\bf B}$ and to the radial direction
in the transverse plane ${\bf k}_\perp$, along which the magnetic
field of the current decreases. The mode-3 photon contributes as
an interaction carrier in the problem of a magneto-static field
produced by a solenoid with its axis along axis 3. In the present
paper, however, we do not consider magneto-static problems.

In the asymptotic limit of high magnetic field $eB\gg k_3^2,~~B\gg
m^2/e\equiv B_0$ the eigenvalue $\kappa_2(0,{\bf k}),$ as
calculated within the one-loop approximation
\cite{batalin,tsai,baier,melrose1,annphys}, with the accuracy of
terms that grow with $B$ only as its logarithm and slower is
\cite{kratkie}
\begin{eqnarray} \kappa_2(0,k_3^2,k_\perp^2) =-\frac{2\alpha
bm^2}{\pi }\exp \left(- \frac{k_\bot^2}{2m^2b}
\right)T\left(\frac{k_3^2}{4m^2}\right),\label{2}\end{eqnarray}
where $b=B/B_0$ and\begin{eqnarray}
T(y)=y\int_{0}^1\frac{(1-\eta^2)\rm d \eta}{1+y(1-\eta^2)}=1-\frac
1{2\sqrt{y(1+y)}}\ln\frac{\sqrt{1+y}+\sqrt{y}}{\sqrt{1+y}-\sqrt{y}}
.\label{T}\end{eqnarray} Note that $0\leq T(y)\leq 1$ for
$y\;\epsilon\; (0,\infty)$. It will be demonstrated in the
subsequent sections that the asymptote $T(y\rightarrow 0)\sim
2y/3$ in (\ref{T}) is responsible for the large-distance
Coulomb-like behavior of the potential in the direction orthogonal
to the field, while the asymptotic value $T(\infty)=1$ introduces
a sort of photon mass and governs the short-range Yukawa-like part
of the potential [see Eqs. (\ref{universal}), (\ref{yuk}) and
(\ref{llargex}) below].

Other eigenvalues, $\kappa_{1,3},$ do not contain the coefficient
$b$ that provides the linear increase of  $\kappa_2$ (\ref{2})
with the magnetic field. Therefore, in the polarization tensor,
whose covariant decomposition is \bee\label{pimunu}\Pi_{\mu
\nu}(k)=\sum_{a=1}^3\kappa_a(k)~\frac{\flat_\mu^{(a)}~
\flat_\nu^{(a)}}{(\flat^{(a)})^2},\eend the components
$\mu,\;\nu=(0,3)$ dominate in the high magnetic field limit in
accord with the idea about the two-dimensioning of the photon
sector.

 Expression for $\kappa_2$ (\ref{2}) is to
be used in (\ref{A0}) or, equivalently, in the expression
\bee\label{bessel}A_0({\bf x})=\frac{q}{2(2\pi)^2}\int_0^\infty
J_0(k_\perp x_\perp)\left(\int_{-\infty}^\infty\frac{\rme^{-\rmi
k_3x_3}\rmd k_3}{
k_\perp^2+k_3^2-\kappa_2(0,k_3^2,k_\perp^2)}\right)\rmd
k_\perp^2\eend obtained from (\ref {A0}) by integration over the
angle between the 2-vectors ${\bf k}_\perp$ and ${\bf x}_\perp$,
which are projections of ${\bf k}$ and ${\bf x}$ onto the plane
transverse to the magnetic field. In (\ref{bessel})
$x_\perp=\sqrt{x_1^2+x_2^2}$ and $J_0$ is the Bessel function of
order zero. We explained in \cite{PRL07} why the $k_3$-integration
here may be extended up to the value $|k_3|=\infty$ in spite of
the limitation on the validity of (\ref{2}) indicated above.

\begin{figure}
\includegraphics[width=0.47\textwidth]{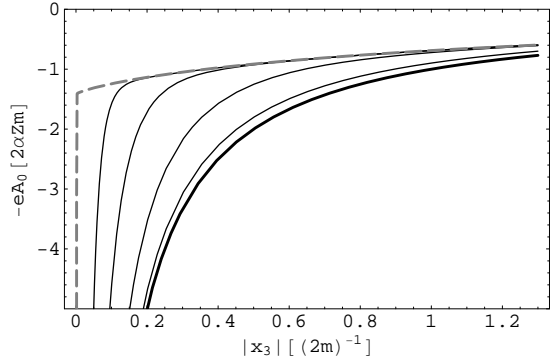}
\caption{\label{fig:1} Electron energy $ -eA_0(x_3,0)$ in the
modified Coulomb potential (\ref{bessel}) of the charge $q=Ze$
plotted along the axis $x_3$ passing through the charge $q$
parallel to the magnetic field, $x_\perp=0$. Thin solid lines
correspond to four values of the magnetic field (from left to
right): $b=10^6,\; b=10^5$, $b=10^4$ and $b=10^3$. Bold solid line
is the Coulomb law $A_0^{\rm C}(x_3,0)=q/(4\pi x_3)$. Thin lines
approach asymptotically the bold line at the both edges of the
figure. Thick dashed broken line corresponds to $b=\infty$. The
abscissa represents the distance in the units $(2m)^{-1}$. The
ordinate represents the potential in the units $2Z\alpha m=Z\times
7.46$ keV.}
\end{figure}

\begin{figure}
\includegraphics[width=0.47\textwidth]{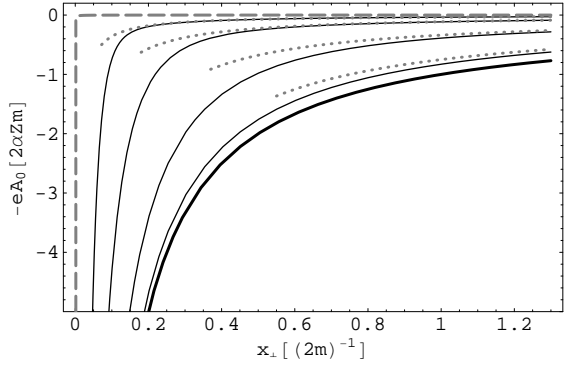}
\caption{\label{fig:2} Electron energy $ -eA_0(0,x_\perp)$ in the
modified Coulomb potential (\ref{bessel}) plotted along the axis
$x_\perp$ passing through the charge $q$ transversely to the
magnetic field, $x_3=0$. Conventions are the same as in Fig.
\ref{fig:1}. Thin lines approach asymptotically the solid line at
the left lower edge of the figure and the short dotted lines
(\ref{llargex}) $A_0(0,x_\perp)=q/(4\pi x_\perp^\prime)$ at the
upper right edge.}
\end{figure}

The results of computer calculation following Eq. (\ref{bessel})
of the shapes of the potential for the two cases, $x_\perp=0$ and
$x_3=0$, are given in Figs. \ref{fig:1} and \ref{fig:2},
respectively \cite{footnote1}. The curves in Figs. \ref{fig:1} and
\ref{fig:2}  manifest the standard Coulomb singularity $1/|{\bf
x}|$, when $|{\bf x}|\rightarrow 0$. Then, they fall up rather
sharply, following a Yukawa-like law within the range of several
Larmour lengths $L_{\rm B}$ to reach the asymptotic long-range
regime that is the Coulomb law $A_0^{\rm C}(x_3,0)= q/4\pi x_3$
for $x_\perp=0$ and $|x_3|\gg m^{-1}$ and $A_0(0,x_\perp)= q/4\pi
x^\prime_\perp$, $x^\prime_\perp=x_\perp \sqrt{1+\alpha b/3\pi}$,
for $x_3=0$ and $x_\perp\gg m^{-1}$ [see Eq. (\ref{llargex})
below]. In what follows we shall be commenting on the features of
the computed curves referring to analytical considerations.

The nontrivial - other than the leading asymptote $\sim k_3^2$
near the point $k_3=k_\perp =0$ - dependence of the polarization
operator eigenvalue (\ref{2}) on the photon momentum components
$k_3$, $k_\perp$ is the spatial dispersion. We shall see in
Section 3 that it is important in the vicinity of the charge,
where the field has a large gradient. As for the anisotropic
behavior far from the charge, to be studied in Section 4, only the
above asymptote is essential, inferable also from the
Heisenberg-Euler Lagrangian.

In Appendix 1 we consider the singular asymptotic behavior of the
potential in the vicinity of the point charge and present its
expansion near $|{\bf x}|=0$. Now we shall consider separately two
additive parts of the potential that decrease by different ways
with increase of the distance ${\bf x}$.

%\section{Short- and long-range parts of the potential}
It is useful to subdivide identically the potential (\ref{bessel})
as:\bee\label{twoparts}A_0({\bf x})= A_{\rm s.r.}({\bf x})+A_{\rm
l.r.}({\bf x})\eend with \bee\label{short}A_{\rm s.r.}({\bf x})
=\frac{q}{2(2\pi)^2}\int_0^\infty J_0(k_\perp
x_\perp)\left[\int_{-\infty}^\infty\frac{\rme^{-\rmi k_3x_3}\rmd
k_3}{ k_\perp^2+k_3^2+{2\alpha b m^2\over\pi}\exp
\left(-{k_\perp^2\over 2m^2b}\right)}\right]\rmd k_\perp^2\eend
and \bee\label{long}A_{\rm l.r.}({\bf
x})=\frac{q}{2(2\pi)^2}\int_0^\infty J_0(k_\perp
x_\perp)\int_{-\infty}^\infty\rme^{-\rmi k_3x_3}\rmd
k_3\nonumber\hspace{5cm}\\\times \left[\frac{1}{
k_\perp^2+k_3^2+\frac{2\alpha bm^2}{\pi }\exp \left(-
\frac{k_\bot^2}{2m^2b} \right)T\left({k_3^2\over
4m^2}\right)}-\frac{1}{k_\perp^2+k_3^2+\frac{2\alpha
bm^2}{\pi}\exp\left(- \frac{k_\bot^2}{2m^2b} \right)}\right]\rmd
k_\perp^2.\eend Eq. (\ref{short}) is the potential (\ref{A0}) or
(\ref{bessel}) taken with the substitution $T\left({k_3^2\over
 4m^2}\right)\Rightarrow T(\infty)=1$ inside $\kappa_2$ (\ref{2}).
We shall see in what follows that $A_{\rm s.r.}({\bf x})$ is a
Yukawa-like potential, singular in the origin, that exponentially
decreases at
 distances of about $\sqrt{\pi/2\alpha}\simeq 15L_{\rm B}$,
 while $A_{\rm l.r.}({\bf x})$ is a finite function that slowly
decreases at large distances (greater than the Compton length
$m^{-1}$) following what will be called anisotropic Coulomb law.
This is the reason why we shall call Eq. (\ref{short}) {\em the
short-range part} and Eq. (\ref{long}) {\em the long-range part}
of the potential.

Consider first the short-range part (\ref{short}) and the limiting
form it takes in the infinite-field limit.

\section{Short-range part}

\subsection{The scaling regime}

\begin{figure}[htb]
  \begin{center}
   \includegraphics[bb = 0 0 405 210,
    scale=1]{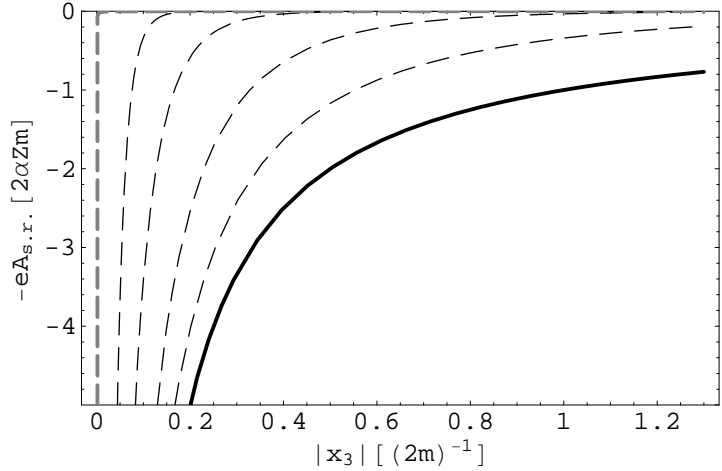}
\caption{%Ratio of the modified Coulomb potential (\ref{bessel}) to
%the standard Coulomb potential (\ref{free2})
%$A_0(0,x_\perp)/A_0^{\rm C}(0,x_\perp)$ drawn as a solid line in
%function of the transverse coordinate $x_\perp$ for the magnetic
%field value $B=10^4B_0$. The dashed line is the constant $(x_\perp
%/x_\perp^\prime)=(1+\alpha B/3\pi B_0)^{-1/2}=0.338$ from the law
%(\ref{llargex}). The dotted line is the Yukawa interpolation. %with $x_3=0$.
%The abscissa is the same as in Fig.\ref{fig:2}
 Electron energy $-eA_{\rm s.r.}(x_3,0)$ in the short-range part
Eq. (\ref{short}) of the potential plotted against the (absolute
value of) longitudinal distance $x_3$ for $x_\perp=0$ (thin dashed
lines
 from left to right  correspond to $b=10^6,\;10^5,\;10^4,\;10^3$). The
 thick
dashed broken line corresponds to $b=\infty$. All the rest is the
same as in Figs \ref{fig:1} and \ref{fig:2}. \label{fig:3}}
  \end{center}
\end{figure}

It is remarkable to note that the short-range part of the
potential (\ref{short}), as measured in the inverse Larmour length
$L_{\rm B}^{-1}=\sqrt{eB}$ units is a universal,
magnetic-field-independent function of coordinates measured in
Larmour units $L_{\rm B}$. To establish this scaling regime let us
make the change of variables in the integral (\ref{short})
$\widetilde{k}_3=k_3L_{\rm B},$ $\widetilde{k}_\perp=k_\perp
L_{\rm B}$ and define the new dimensionless coordinates
$x_3=\widetilde{x}_3L_{\rm B},$ $x_\perp=\widetilde{x}_\perp
L_{\rm B}.$ %After that, we may substitute unity in place of the
%function $T(\widetilde{k}_3^2/4m^2L_{\rm B}^2)$, since it is the
%value of this bounded function in the limit
%$(eB/4m^2)=1/4m^2L_{\rm B}^2=\infty$.
Then Eq. (\ref{short})
becomes\bee\label{difference3}\hspace{-3.5cm}A_{\rm s.r.}({\bf
x})=\frac {q}{2(2\pi)^2L_{\rm B} }\int_0^\infty
J_0(\widetilde{k}_\perp\widetilde{
x}_\perp)\int_{-\infty}^\infty\frac{\rme^{-\rmi
\widetilde{k}_3\widetilde{x}_3}\rmd \widetilde{k}_3\rmd
\widetilde{k}_\perp^2}{
\widetilde{k}_\perp^2+\widetilde{k}_3^2+\frac{2\alpha}{\pi}\exp
\left(-\frac{\widetilde{k}^2_\perp}2\right)}.\eend This is an even
function of $x_3$. The $\widetilde{k}_3$-integration here can be
performed by calculating the residues in the poles on the
imaginary axis in the points \bee\label{neweqktilde}
\widetilde{k}_3=\pm\rmi
\sqrt{\widetilde{k}_\perp^2+\frac{2\alpha}{\pi}\exp
\left(-\frac{\widetilde{k}^2_\perp}2\right)}\eend with the upper
sign taken for $x_3>0$ and the lower one for $x_3<0$. Finally one
gets \bee\label{universal}\hspace{-1.5cm} A_{\rm s.r}({\bf x})=
\frac{\widetilde{A}(\widetilde{\bf x})}{L_{\rm B}}= \frac {q}{8\pi
L_{\rm B}}\int_0^\infty J_0(\widetilde{k}_\perp\widetilde{
x}_\perp)\frac{\rme^{-|\widetilde{x}_3|\sqrt{\widetilde{k}_\perp^2
+\frac{2\alpha}\pi\exp\left(-\frac{\widetilde{k}_\perp^2}2\right)}}}
{\sqrt{\widetilde{k}_\perp^2+\frac{2\alpha}{\pi}\exp
\left(-\frac{\widetilde{k}^2_\perp}2\right)}}\rmd
\widetilde{k}^2_\perp.\eend Here the universal function
$\widetilde{A}(\widetilde{\bf x})$ depends on the magnetic field
 through its spatial arguments $\widetilde{\bf x}$ only.
 Eq. (\ref{short}) (or (\ref{difference3}) and (\ref{universal})) is
 illustrated in Fig. \ref{fig:3} drawn for $x_\perp =0$ by a
 computer.
The simple representation (\ref{universal}) can be further
simplified if $x_3$ or $x_\perp$ are large in the Larmour scale:
$|\widetilde{x}_3|\gg 1,$ or $|\widetilde{x}_\perp|\gg 1.$ In this
case the integration in (\ref{universal}) is restricted to the
domain $\widetilde{k}_\perp^2\ll 1$ where the exponential $\exp
(-\widetilde{k}_\perp^2/2)$ should be taken as unity. Then
(\ref{universal}) is reduced to the Yukawa law  for the
short-range part of the potential
\bee\label{yuk}\hspace{-2cm}A_{\rm s.r.}({\bf x})\simeq A_0^{\rm
Y}({\bf x})=\frac {q}{4\pi L_{\rm B}}\frac
{\exp\,[-(2\alpha/\pi)^{1/2}
{\sqrt{\widetilde{x}_\perp^2+\widetilde{x}_3^2}}
\,]}{\sqrt{\widetilde{x}_\perp^2+\widetilde{x}_3^2}}=\frac
q{4\pi}\frac{\exp\,[-(2\alpha b/\pi)^{1/2}m|{\bf x}|\,]} {|{\bf
x}|}.\eend It reflects the Debye screening of the charge by the
polarized vacuum. Eq. (\ref{yuk}) can be established if we return
to the previous representation (\ref{difference3}), which can then
be traced back to (\ref{bessel}) with
\bee\label{mass}-\kappa_2(0,\infty,0)=\frac{2\alpha}{\pi L_{\rm
B}^2}T(\infty)=m^2\frac{2\alpha b}\pi\equiv M^2\eend substituted
for $-\kappa_2(0,k_3^2, k_\perp^2)$ in the denominator. Here $M$
is the "effective photon mass" $\;$of Ref. \cite{kuznets}. Write
it as \bee\label{mass2} M=\frac e{\pi\sqrt{2}\;L_{\rm B}}.\eend
The deviation of (\ref{universal}) from (\ref{yuk}) when
$\widetilde{x}_\perp$ and $\widetilde{x}_3$ are both small is not
very important against the background of the singularity of the
short-range part of the potential $A_{\rm s.r.}({\bf x})$ near the
charge provided by the divergency of (\ref{short}) in the origin
$\bf x$=0. (See Appendix 1 for the asymptotic expansion of the
potential near the origin). Therefore, the Yukawa law (\ref{yuk})
is approximately fulfilled also in the close vicinity of the
charge. Correspondingly, the curves in Fig. \ref{fig:3} could not
be distinguished from the Yukawa law (\ref{yuk}) in the scale of
the figure.

 Eqs. (\ref{universal}) and (\ref{yuk}) establish the short-range character of the
static electromagnetic forces in the Larmour scale. The
corresponding effective mass (\ref{mass}) coincides with the
topological photon mass $e_{\rm Sch}/\sqrt{\pi}$ in the
2-dimensional Schwinger electrodynamics \cite{schwinger} provided
that the dimensional  fermion charge $e_{\rm Sch}$ of that theory
is identified as $e_{\rm Sch}=\sqrt{\pi/2}\;e/L_{\rm B}$. Stress
that the zero photon mass understood as its rest energy is also
present as a consequence of the gauge invariance reflected in the
approximation-independent relation $\kappa_a(0,0,0)=0.$
Correspondingly, the potential, produced by a static charge,
should be long-range for sufficiently large distances. This is the
case, indeed. The long-range character of the electromagnetic
interaction is restored at the distances of the Compton scale, as
we shall see in the next Section. The carrier of the long-ranged
interaction  will be $A_{\rm l.r.}({\bf x})$ (\ref{long}). The
Debye screening obtained here in the vacuum completely depends on
the fact that the function (\ref{T}) tends to unity for large
longitudinal momentum $k_3\to\infty$, i.e. on spatial dispersion.
In this point the situation is different from the case of a
medium, where the Debye screening can be achieved \cite{zeitlin}
in expressions, obtained from the thermodynamical potential, which
is the analog of the effective  Lagrangian for that case. The
difference with the medium  is also in that the long-range part of
the $static$ potential is absent in that case in spite of the
gauge invariance, because it implies that the appropriate
polarization tensor components should disappear in the long-wave
limit $k_0=0$, $\bf k$=0 only if $\bf k$ is set equal to zero
first \cite{fradkin}, thus providing the zero value to the photon
$magnetic$ mass.

\subsection{The limiting $b=\infty$ form}
The short-range part (\ref{short}) and (\ref{difference3}) tends to
zero, when $b\rightarrow\infty$ for any nonzero distance, ${\bf
x}\neq 0$, from the charge. This follows from the fact that
$\widetilde{A}(\widetilde{x})$, defined in (\ref{universal}), tends
to zero with the exponential speed (\ref{yuk}) when
$\widetilde{x}_3=x_3/L_{\rm B}$ or $\widetilde{x}_\perp=x_\perp
/L_{\rm B}$ tends to infinity. As the magnetic field $b$ grows more
and more, the curves representing the potential (\ref{universal})
for the special case of $x_\perp=0$ in Fig. \ref{fig:3} stick closer
and closer to the vertical axis, the spacing between the curves and
this axis becoming infinitely thin in the limit $b=\infty$. The area
$(q/2\pi)S$ of the region restricted by any curve (\ref{universal})
and the $x_3$-axis in the domain $-\infty<x_3\leq -L_{\rm B},\;
L_{\rm B}\leq x_3<\infty$ \bee\label{area2}\frac
{q}{2\pi}S=2\int^\infty_{L_{\rm B}} A_{\rm s.r.}(x_3,0)\rmd
x_3=2\int^\infty_1 \widetilde{A}(\widetilde{x}_3,0)\rmd
\widetilde{x}_3=\frac
q{4\pi}\int_0^\infty\frac{\rme^{-\sqrt{\widetilde{k}_\perp^2
+\frac{2\alpha}\pi\exp\left(-\frac{\widetilde{k}_\perp^2}2\right)}}
\rmd\widetilde{k}^2_\perp}{\widetilde{k}_\perp^2
+\frac{2\alpha}\pi\exp\left(-\frac{\widetilde{k}_\perp^2}2\right)}\eend
is a finite number, with $S=2.180$, independent of the magnetic
field. If the Yukawa approximation (\ref{yuk}) is used in
(\ref{area2}) in place of (\ref{universal}), approximately the same
numerical value for $S$ is achieved: $S=-{\rm
Ei}(-\sqrt{2\alpha/\pi})\simeq 2.176$, where \bee{\rm
Ei}(u)=-\int_{-u}^\infty\exp(-y)y^{-1}\rmd y\eend is the exponential
integral. In the limit $b=\infty$, $L_{\rm B}=0$ the width of the
strip $|x_3|\leq L_{\rm B}$ excluded from the integration in Eq.
(\ref{area2}) is zero, and the latter becomes the whole area above
the limiting potential. Thus, in the infinite-magnetic-field limit
the short-range part (\ref{short}) of the potential taken on the
axis drawn through the point charge $q$ along the magnetic field
direction becomes the $\delta$-function:\bee\label{deltax3}
\left.A_{\rm s.r.}(x_3,0)\right|_{b=\infty}=2.180\frac
q{2\pi}\delta(x_3).\eend The limiting $\delta$-function here is
understood in the following sense. Given a test-function $t(x_3)$,
one has\bee\label{deltalimit}
\lim_{b\rightarrow\infty}\left[\int^{-L_{\rm B}}_{-\infty} A_{\rm
s.r.}(x_3,0)t(x_3)\rmd x_3+\int_{L_{\rm B}}^\infty A_{\rm
s.r.}(x_3,0)t(x_3)\rmd x_3\right]=2.180\frac q{2\pi}t(0).\eend [This
equation directly follows from the scaling law, the first equality
in (\ref{universal})]. In this sense it will be used in Section IV
and Appendix 2, where we shall see that the $\delta$-singularity of
the potential (\ref{deltax3}) leads only to a finite contribution to
the atomic ground-state energy in an infinite magnetic field in
contrast to the contribution of the primary Coulomb potential.

 Analogously, we may write a
$\delta$-function for the limiting form of the short-range part of
the potential along any direction $x$, $|x|=x_\perp$, in the plane
orthogonal to the magnetic field containing the point charge $q$. In
place of (\ref{area2}) one has\bee\label{area3}\frac
{q}{2\pi}S_\perp=2\int^\infty_{L_{\rm B}} A_{\rm s.r.}(0,x)\rmd
x=2\int^\infty_1 \widetilde{A}(0,\widetilde{x})\rmd
\widetilde{x}=\nonumber\\=\frac q{2\pi}\int_0^\infty\frac{
 1-\widetilde{k}_\perp
J_0(\widetilde{k}_\perp)+\frac {\pi\widetilde{k}_\perp}{2}[
J_0(\widetilde{k}_\perp){\bf H_1}(\widetilde{k}_\perp)-
J_1(\widetilde{k}_\perp){\bf H_0}(\widetilde{k}_\perp)]
}{\sqrt{\widetilde{k}_\perp^2
+\frac{2\alpha}\pi\exp\left(-\frac{\widetilde{k}_\perp^2}2\right)}}
\rmd\widetilde{k}_\perp=\frac q{2\pi}2.178. \quad\eend Here $
J_{0,1}$ and ${\bf H_{0,1}}$ are, resp., the Bessel and Struve
functions of orders zero and one. We used the integral 6.512.8 and
the representation 8.551.1 in the reference book \cite{ryzhik} for
calculating (\ref{area3}). Note that the Struve functions at large
argument  decrease and oscillate asymptotically like the Neuman
functions; besides ${\bf H}_1$ includes a constant asymptotic term
$2/\pi$. The integral (\ref{area3}) converges: the divergence
caused by the unity term in the nominator is cancelled by the two
products of two oscillating asymptotic terms in the square
brackets.

From (\ref{area3}) and the fact that for $x_\perp\neq 0$ the
short-range potential (\ref{universal}) disappears in the
$b\to\infty$ limit we have\bee\label{deltaxperp} \left.A_{\rm
s.r.}(0,x)\right|_{b=\infty}=2.178\frac q{2\pi}\delta(x).\eend The
different coefficients in (\ref{deltax3}) and (\ref{deltaxperp})
manifest the anisotropy. Note that the Coulomb singularity of the
(short-range part of) the potential in the origin $q/4\pi|{\bf x}|$
is isotropic \cite{to avoid}.

\section{Long-range part}

We have finished the consideration of the short-range part and
will proceed with considering the long-range part $A_{\rm
l.r.}({\bf x})$ (\ref{long}). % Curves drawn following Eq.
%(\ref{long}) are presented in FIG. \ref{fig:5}.
Simplifying
expressions will be obtained for large-distance behavior in
Subsection A and for the long-range part taken on the axis
$x_\perp=0$ in Subsection B. We shall see in Subsection B that in
the limit $b=\infty$ the long-range part, as well as the whole
potential, is concentrated on this axis, making an infinitely thin
tube or string. We shall study the potential along the string in
more detail in Subsection B.
\subsection{Long-distance behavior of the long-range part $A_{\rm l.r.}({\bf x})$}
Once we have seen in the previous Section that the short-range
part  $A_{\rm s.r.}({\bf x})$ is as a matter of fact concentrated
within the region of a few $L_{\rm B}$, for larger distances,
$|{\bf x}|\gtrsim m^{-1},$ the whole potential (\ref{bessel}) and
its long-range part (\ref{long}) are the same. For this reason in
this Subsection we shall deal directly with (\ref{bessel}).

\subsubsection{Large $x_\perp$ in Larmour scale}

For large transverse distances the term linearly growing with the
magnetic field (\ref{2}) %dominates in the vacuum polarization, it
leads to suppression of the static potential in the transverse
direction.

To be more precise, consider the region
\bee\label{largexperp}x_\perp\gg \frac
{m^{-1}}{\sqrt{2b}}=\frac{L_{\rm B}}{\sqrt{2}}.\eend Once the
Bessel function $J_0$ in (\ref{bessel}) oscillates and decreases
for large values of its argument $k_\perp x_\perp$, the main
contribution into the integral over $k_\perp^2$ in (\ref{bessel})
comes from the integration variable domain $k_\perp^2\ll 2m^2b$,
and the dependence upon $k_\perp^2$ in $\kappa_2$ may thus be
disregarded. Then the $k_\perp^2$-integration in (\ref{bessel})
can be explicitly performed to give (we use Eq.6.532.4 of the
reference book \cite{ryzhik})
\bee\label{largexperp2}A_0(x_3,x_\perp)\simeq \frac {2q}{(2\pi
)^2}\int_0^{\infty}
\mathcal{K}_0\left(x_\perp\sqrt{k_3^2-\kappa_2(0,k_3^2,0)}\right)\cos
(k_3x_3)\rmd k_3,\eend where $\mathcal{K}_0$ is the McDonald
function of order zero, and
\bee\label{kappa00}\kappa_2(0,k_3^2,0)=-\frac{2\alpha b}\pi
m^2T\left(\frac {k_3^2}{4m^2}\right).\eend %We have now to consider
%regions of small and large $|x_3|$.
As the McDonald function $\mathcal{K}_0$ decreases exponentially
when its argument increases, only small values of the square root
contribute into integral (\ref{largexperp2}), %which implies that
%Eq.(\ref{T(0)}) should be again used to lead to (\ref{llargex}).
%Due to the oscillating character of cos$(k_3x_3)$ in
%\ref{largexperp2}), for $|x_3|$ as large as
%\bee\label{largex3}|x_3|\gg \frac {m^{-1}}{2}\eend
and the $k_3$-integration domain in it is restricted to the
interval $k_3^2\ll 4m^2$, wherein \bee\label{T(0)}
T\left(\frac{k_3^2}{4m^2}\right)\simeq \frac{k_3^2}{6m^2}.\eend
Then the potential form (\ref{largexperp2}) becomes (we use Eq.
6.671.14 of the reference book \cite{ryzhik})
\bee\label{llargex}A_0(x_3,x_\perp)\simeq \frac
{2q}{(2\pi)^2}\int_0^\infty\mathcal{K}_0\left(x_\perp
k_3\sqrt{1+\frac{\alpha b}{3\pi}}\right)\cos (k_3x_3)\rmd
k_3\nonumber\\=\frac 1{4\pi}\frac
{q}{\sqrt{(x^\prime_\perp)^2+x_3^2}},\quad\qquad
x^\prime_\perp=x_\perp\left(1+\frac{\alpha
b}{3\pi}\right)^{1/2},~~x^\prime_\perp>x_\perp.\eend % We have
%$derived Eq. (\ref{llargex}) in the space region specified by the
%two conditions (\ref{largexperp}), (\ref{largex3}).

Eq. (\ref{llargex}) is an anisotropic Coulomb law, according to
which the attraction force decreases with distance from the source
along the transverse direction faster than along the magnetic field
(remind that $b\equiv (B/B_0)\gg 1$), but remains long-range. The
equipotential surface is an ellipsoid stretched along the magnetic
field. The electric field of the charge  ${\bf E}=-{\bf
\nabla}A_0(x_3,x_\perp),$ as written in Cartesian components, is the
vector $(q/2\pi)(x_3^2+\beta^2x_\perp^2)^{-3/2}~(\beta^2 x_1
~~\beta^2 x_2 ~~x_3)$, where $\beta=(1+\alpha b/3\pi)^{1/2}$. It is
not directed towards the charge, but makes an angle $\phi$ with the
radius-vector $\bf r$,
cos$\phi=(x_3^2+\beta^2x_\perp^2)(x_3^2+\beta^4x_\perp^2)^{-1/2}
(x_3^2+x_\perp^2)^{-1/2}.$ If $x_\perp\neq 0$, in the limit of
infinite magnetic field, $\beta\to\infty$, the electric field of the
point charge is directed normally to the axis $x_3$, since the ratio
$(E_3/E_\perp )\to 0$ (although $E_3$ and $E_\perp$ are both equal
to zero in this limit outside the string). But if $x_\perp= 0$, the
electric field is directed along the external magnetic field. It
looks like the electric field compresses the string. This regime
corresponds to the dielectric permeability of the vacuum independent
of the frequency, with its dependence on $\bf k$ (spatial
dispersion) being reduced solely to that upon the angle in the space
($cf,$
\cite{zhetf}). %According to \cite{zhetf} the transverse refraction
%index for the mode-2 photons is just \bee
%n_2^\bot=\left(1+\frac{\alpha b }{3\pi}\right)^\frac
%1{2}=\left(1+7.7\cdot 10^{-4}\frac B{B_{cr}}\right)^{\frac 1{2}}.
%\label{13} \eend
%\subsection{$x_\perp$ large in Compton scale, $x_3$ any}
%We may remove the condition (\ref{largex3}), if we pass to the
%transverse distances still larger than the ones limited by the
%condition (\ref{largexperp}), and state now that the anisotropic
%Coulomb law (\ref{llargex}) also holds true, provided that
%\bee\label{vlargexperp}x_\perp\gg\frac{m^{-1}}{2},\qquad {\rm
%with~~ any}~~ |x_3|.\eend To see this, note that

%A numerical computation of the potential (\ref{A0}),
%5(\ref{bessel}) as a function of $x_\perp$ at a fixed $x_3=0$ was
%done for $B=10^4B_0$ ($m^{-1}=100 L_{\rm B}$ with this choice of
%$b$).
% The results presented in Figures 1,2,3 cover  together the
%interval extending from the transverse distances as small as
%$x_\perp\simeq 10^{-4}m^{-1}\approx 10^{-2}L_{\rm B}$ and up to
%the distances as large as $x_\perp\simeq 10^{2}m^{-1}$.
 The result (\ref{llargex}) is in
agreement with  the  curves in Fig. \ref{fig:2} in the large
$x_\perp$ domain.
%\ref{fig:4} plotted against $x_\perp$. %Really,reaches the asymptote
%$A_0(0,x_\perp)\simeq q/4\pi x^\prime_\perp$, $x^\prime_\perp
%=2.957 x_\perp$ already for the transverse distance $x_\perp\geq
%1.5m^{-1}$. In the interval $L_{\rm B}<x_\perp < m^{-1}$, wherein
%also the expression (\ref{largexperp2}) with $x_3=0$
%\bee\label{largexperp3}A_0(0,x_\perp)\simeq \frac
%{q}{2\pi^2}\int_0^{\infty}
%\mathcal{K}_0\left(x_\perp\sqrt{k_3^2-\kappa_2(0,k_3^2,0)}\right)\rmd
%k_3,\eend holds true,
%The curves in Fig.\ref{fig:2}, approach the standard Coulomb
%law when $x_\perp\rightarrow 0$ %at the left edge $x_\perp\asymp L_{\rm
%B}=10^{-2}m^{-1}$
%in accordance with (\ref{expand}), as explained above,  rather
%sharply fall up, following the Yukawa law (\ref{yuk}) in the
%Larmour range $x_\perp\sim L_{\rm B}$ to  reach the asymptotic
%long-range regime $A_0(0,x_\perp)\simeq q/4\pi x^\prime_\perp$ for
%larger $x_\perp$ in the region (\ref{largexperp}).
%% Therefore, in the
%interval $L_{\rm B}<x_\perp < m^{-1}$ the potential behaves itself
%as short-range, although later, for $x_\perp\gg m^{-1}$ it
%approaches the long-range tail (\ref{llargex}).
\subsubsection{Large $x_3$}

It remains to consider the remote coordinate region of large
$x_3$, complementary to ({\ref{largexperp}).

To this end we apply the residue method to the inner integral over
$k_3$ in (\ref{bessel}).
 Using the
integral representation (\ref{T})  the function $\kappa_2$
(\ref{2}) may be, for a fixed positive value of $k_\perp^2$,
analytically continued from the real values of the variable  $k_3$
into the whole complex plane of it, cut along two fragments of the
imaginary axis. In the lower half-plane the cut runs from Im$~
k_3=-2m$ down to Im $k_3= -\infty$, while in the upper half-plane
it extends within the limits $2m \leq {\rm Im}~ k_3\leq\infty$.
Other singularities of the $k_3$-integrand in (\ref{bessel}) are
poles yielded by zeros of the denominator, i.e. solutions of the
equation (associated with the photon dispersion
equation)\bee\label{dispersion}
k_\perp^2+k_3^2-\kappa_2(0,k_3^2,k_\perp^2)=0.\eend  As $k_\perp$
varies within the limits $(0,\infty)$ two roots of this equation
$k^\pm_3=\pm \rmi K(k_\perp)$ move along the imaginary axis from
the point $K(0)=0$ to the points  $k^\pm_3=\pm \rmi K(\infty)=\pm
\rmi 2m$ \cite{annphys,nuovcimlet,zhetf}. There is yet another
branch of the solution to equation (\ref{dispersion}),
corresponding to the photon absorption via the $\gamma\rightarrow
e^+e^-$-decay, but the corresponding poles lie in the nonphysical
sheet of the described complex plane, behind the cuts, and will
not be of importance for the consideration below.

Let us consider positive values of $x_3$. Negative values can be
handled in an analogous way. Turning the positive part of
integration path $0\leq k_3\leq\infty$ clockwise to the lower
half-plane by the angle $\pi/2$, and the negative part $-\infty\leq
k_3\leq 0$ counterclockwise by the same angle, and referring to the
fact that the exponential $\exp (-\rmi k_3x_3)$ decreases, for
$x_3>0$, in the lower half-plane of $k_3$ as
$|k_3|\rightarrow\infty$ so that the integrals over the remote arcs
may be omitted, we get a representation for the inner integral in
(\ref{bessel})\bee\label{inner}\int_{-\infty}^\infty\frac{\rme^{-\rmi
k_3x_3}\rmd k_3}{
k_\perp^2+k_3^2-\kappa_2(0,k_3^2,k_\perp^2)}\nonumber\\=\rmi
\int_{2m}^\infty\rme^{-|k_3|x_3} \Delta(|k_3|^2,k_\perp^2)\,\rmd
|k_3| - \rmi\, 2 \pi \exp\,[-K(k_\perp^2)x_3\,]\,{\rm
Res}(k_\perp^2)\,,\eend where Res$(k_\perp^2)$ designates the
residue of the expression $D_2 (0,-|k_3|^2,k_\perp^2)=\left(
k_\perp^2-|k_3|^2-\kappa_2(0,-|k_3^2|,k_\perp^2)\right)^{-1}$ in the
pole $k_3^-=-iK(k^2_\perp)$, while
$\Delta(|k_3|^2,k_\perp^2)=D_2(0,-|k_3|^2+\rmi 0,k_\perp^2)-
D_2(0,-|k_3|^2-\rmi 0,k_\perp^2)$ is the cut discontinuity. It was
explained above that $0<K(k_\perp^2)< 2m$ everywhere but in the
limit $k_\perp\rightarrow\infty$, where $K=2m$. Consequently the
residue term in (\ref{inner}) dominates over the cut-discontinuity
term everywhere in the $k_\perp$-integration domain in the outer
integral in (\ref{bessel}), except for the region near $k_\perp=
\infty$. In this limit, however, $\kappa_2$ disappears due to the
exponential in (\ref{2}), together with the cut discontinuity, since
the latter is only due to the branching points in the function
(\ref{T}). Therefore, keeping the residue term in (\ref{inner}) as
the leading one, we neglect the contribution that decreases with
large longitudinal distance at least as fast as $\exp (-2m|x_3|).$
In this way we come to the asymptotic representation of the
potential (\ref{bessel}) in the region of large longitudinal
distances $|x_3|\gg (2m)^{-1}$ (negative values of $x_3$ at this
step are also included - to handle them one should rotate the
fragments of the integration path in the directions opposite to the
above) \bee\label{residue}A_0({\bf x})\simeq \frac
{q}{8\pi}\int_0^\infty \frac{J_0(k_\perp
x_\perp)\,\exp\,[-K(k^2_\perp)|x_3|]\,\rmd
k^2_\perp}{K(k^2_\perp)[1+H(-K^2(k^2_\perp),k_\perp^2)]},\eend where
\bee\label{H} H(k_3^2,k_\perp^2)=\frac{2\alpha bm^2}\pi\exp
\left(-\frac{k_\perp^2}{2m^2b}\right)\frac\rmd{\rmd k_3^2}
T\left(\frac{k_3^2}{4m^2}\right). \eend Here $K^2(k_\perp^2)$ is the
solution of equation (\ref{dispersion}) in the negative region of
the variable $k_3^2$ - see \cite{zhetf} for its form. $K(\infty)=
2m$, $K(0)=0$. $T$ is given by (\ref{T}).

Due to the exponential factor in the integrand of (\ref{residue}),
for large $x_3$ the main contribution  comes from the integration
region of $k_\perp$ that provides minimum to the function
$K(k_\perp)$. The minimum value of $K(k_\perp)$ is  zero. It is
achieved in the point $k_\perp=0$ - a manifestation of the fact that
the photon mass defined as its rest energy is strictly equal to zero
owing to the gauge invariance: $\kappa_a(k_0=k_3={\bf k}=0)=0$. In
view of (\ref{kappa00}) and (\ref{T(0)}), near the point $k_\perp=0$
the dispersion equation (\ref{dispersion}) has the solution
$K(k_\perp)=k_\perp/\sqrt{1+\alpha b/3\pi}$. Simultaneously, near
the minimum point $1+H(0,0)=1+\alpha b/3\pi$. With these
substitutions and the use of 6.611.1 of \cite{ryzhik}, Eq.
(\ref{residue}) becomes again the anisotropic Coulomb law
(\ref{llargex}) $(q/4\pi)/[(x^\prime_\perp)^2+x_3^2]^{1/2}$. We
have, therefore, established its validity everywhere in the region
remote from the center, irrespectively of the direction.

In agreement with this result the curves in Fig. \ref{fig:1} for
$A_0(x_3,0)$ approach the Coulomb law $q/4\pi |x_3|$ as $|x_3|$
grows.
\begin{figure}[htb]
\begin{center}
  \includegraphics[bb = 0 0 405 210,
  scale=1]{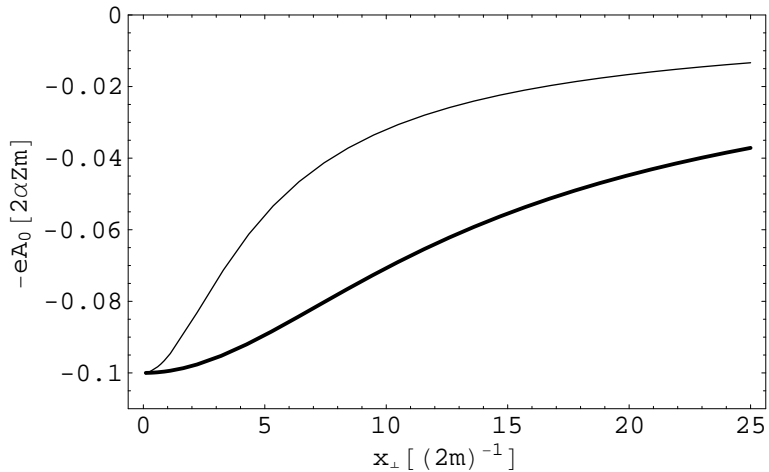}
\caption{ Electron energy $-eA_0(x_3,x_\perp)$ in the modified
Coulomb potential (\ref{residue}) with $q=Ze$ plotted against the
transverse coordinate $x_\perp$ with  the longitudinal coordinate
fixed at the large value $x_3=10(2m)^{-1}$. Thin solid line
corresponds to the magnetic field  value $B=10^4B_0$. Bold solid
line is the standard Coulomb law (\ref{free2}) $-eA_0^{\rm
C}(x_3,x_\perp)= - 2\alpha Z m[(2mx_\perp)^2+100]^{-1/2}$. The
thin line is indistinguishable from the anisotropic Coulomb law
(\ref{llargex}) in the scale of the drawing.  The coordinate axes
are the same as in Fig. \ref{fig:2}} \label{fig:4}
 \end{center}
\end{figure}
%\begin{figure}[htb]
 % \begin{center}
  % \includegraphics[bb = 0 0 405 210,
   % scale=1]{fig5.EPS}
%\caption{The same as Fig.\ref{fig:4}, but viewed at a detailed
%scale near $x_\perp=0$. The thin solid line is the anisotropic
%Coulomb law (\ref{llargex}) $qm/2\pi ((0.338\cdot
%2mx_\perp)^2+100)^{-1/2}$} \label{fig:5}
 % \end{center}
%\end{figure}
The difference %(\ref{difference2})
between the potential $A_0(x_3,0)$ and its large-$x_3$ asymptote
$q/4\pi|x_3|$ decreases in Fig. \ref{fig:1} at least as fast as
exp$(-2m|x_3|)$ (see \cite{arxive} for the derivation of the
latter statement).

Eq. (\ref{residue}) was used for computer calculation with the
large value $x_3=10m^{-1}$. It has  led to the curve shown in
%figures labelled as 4 and 5 in Ref. \cite{arxive}.
Fig. \ref{fig:4}. %, \ref{fig:5}.
In the region  (\ref{largexperp})
it agrees %(\ref{shortperp}), valid for small $x_\perp$ (\ref{xperp}) and any
%$x_3$, and
with the result (\ref{llargex}), valid  in that region %for large $x_\perp$
%(\ref{largexperp}) %and large $x_3$ (\ref{largex3}).
[$L_{\rm B}=0.02 (2m)^{-1}$ for $b=10^4$]. In practice
(\ref{residue}) and (\ref{llargex}) are the same. A small
deviation of the potential curve $A_0(10/2m,x_\perp)$ from
(\ref{llargex}) may be seen in  Fig. 5 of Ref. \cite{arxive},
drawn in a more detailed scale for small $x_\perp$.

 \subsection{The long-range part on the axis $x_\perp=0$ and its limiting form for $b=\infty$}

Curves drawn for $A_{\rm l.r.}(x_3,0)$ by a computer following Eq.
(\ref{long}) are presented in Fig. \ref{fig:5}.

\begin{figure}[htb]
  \begin{center}
   \includegraphics[bb = 0 0 405 210,
    scale=1]{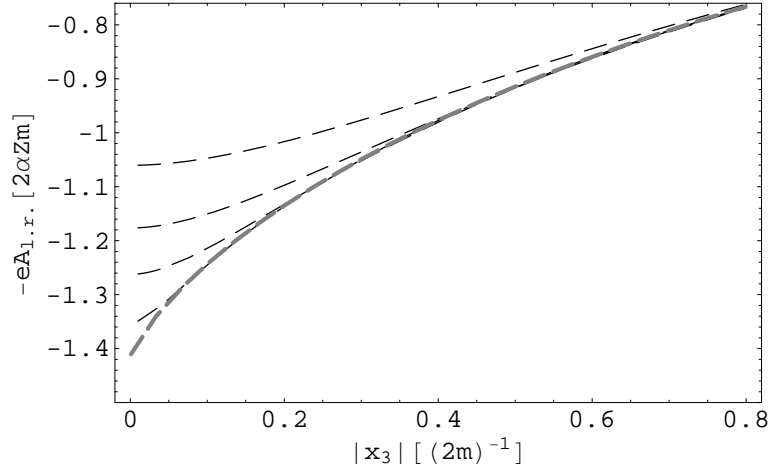} \caption{Electron energy $-eA_{\rm l.r.}(x_3,0)$ in the
    long-range part (\ref{long}) of the potential for $b=\infty$ (dashed thick line) and $b=10^6,\;
    10^5$, $3\times 10^4$, $10^4$ (dashed lines from bottom to top)} %Potential curves  for $b=10^4, 10^5,
%10^6$ approaching their long-range asymptotes}
\label{fig:5}
  \end{center}
\end{figure}
Here we study the form the long-range part (\ref{long}) of the
potential takes in the limit $b=\infty$. %for all distances from the
%charge.

%Unlike Eq.(\ref{short}) whose limit was the $\delta$-function, the
%long-range part (\ref{long}) has a finite limit as the magnetic
%field tends to infinity. The reason is in that the double integral
%(\ref{long}) converges everywhere (the origin included) for every
%value of the magnetic field.
First consider the  case $x_\perp\neq 0,\;x_3\neq 0$. As we saw in
Subsection B of Section 3  the short-range part of the potential
tends in this case to zero as $b\rightarrow\infty$. Therefore, the
limits of the whole potential and of its long-range part are the
same. For this reason to achieve the claimed goal it is sufficient
to consider the limit of (\ref{bessel}). After the change of the
integration variable $k_\perp=\widetilde{k}_\perp m\sqrt{b}\;$ Eq.
(\ref{bessel}) becomes\bee\label{long1} A_{0}({\bf
x})=\frac{q}{2(2\pi)^2}\int_0^\infty J_0(\widetilde{k}_\perp
m\sqrt{b}x_\perp)\int_{-\infty}^\infty\frac{\rme^{-\rmi
k_3x_3}\rmd k_3\rmd \widetilde{k}_\perp^2}
{\widetilde{k}_\perp^2+{k_3^2\over m^2b}+\frac{2\alpha}{\pi }\exp
\left(- \frac{\widetilde{k}_\bot^2}{2} \right)T\left({k_3^2\over
4m^2}\right)}%-\frac{1}{\widetilde{k}_\perp^2+{k_3^2\over m^2b}
%+\frac{2\alpha}{\pi}\exp\left(- \frac{\widetilde{k}_\bot^2}{2}
%\right)}\right\}
\eend When $b\gg (\pi/2\alpha)$ one can disregard the ratio
${k_3^2\over m^2b}$ in the denominator.

For any finite $x_\perp$ the argument of the Bessel function in
(\ref{long1}) is large, therefore we may use the same procedure as
the one that led us from (\ref{bessel}) to (\ref{largexperp2}) and
(\ref{llargex}). Then we obtain \bee\label{lim} \left.A_{0}
(x_3,x_\perp)\right|_{b\rightarrow\infty,\;x_\perp\neq 0}= \frac
{2q}{(2\pi )^2}\int_0^{\infty}\mathcal{K}_0\left(x_\perp k_3
\sqrt{\frac{\alpha b}{3\pi} }\right)%-\mathcal{K}_0\left(x_\perp m
%\sqrt{\frac{2\alpha b}{\pi} }\right)\right]
\cos (k_3x_3)\rmd k_3\nonumber\\={1\over 4\pi}\frac
q{\sqrt{x_\perp^2(\alpha b/3\pi)+x_3^2}}\simeq
{qm\over4\sqrt{\alpha\pi/3}}{L_{\rm B}\over x_\perp}\rightarrow 0
.\eend This means that outside the $x_3$-axis %(when also $x_3\neq 0)$
the potential (\ref{bessel}) turns to zero as the ratio
$L_{\rm B}/x_\perp$. Since its short-range part (\ref{short}) or
(\ref{yuk}) decreases with $b$ exponentially, the  result
(\ref{lim}) holds for the long-range part (\ref{long}) as well.

 The situation is different on the axis $x_\perp=0$.
 By making in Eq. (\ref{long}) the same change of the variable
$k_\perp$ as above
    and, again, neglecting $k_3^2/4m^2b$ in the denominators
   we come to the limiting ($b=\infty$) form of the long-range part of the potential $x_\perp=0$,
   independent of the magnetic field\bee\label{envelope}\left.A_{\rm
   l.r.}(x_3,0
)\right|_{b=\infty}=\frac{q}{(2\pi)^2}\int_0^\infty
\int_0^\infty\cos (k_3x_3)\rmd k_3\nonumber\hspace{5cm}\\\times
\left\{\frac{1}{ \widetilde{k}_\perp^2+\frac{2\alpha}{\pi }\exp
\left(- \frac{\widetilde{k}_\perp^2}{2} \right)T\left({k_3^2\over
4m^2}\right)}-\frac{1}{\widetilde{k}_\perp^2+\frac{2\alpha
}{\pi}\exp\left(- \frac{\widetilde{k}_\perp^2}{2}
\right)}\right\}\rmd \widetilde{k}_\perp^2.\eend

This is the analytic representation of the envelope curve in Fig.
\ref{fig:1}. To understand this, note that the overall potential
is the sum of the short- and long-range parts, according to
(\ref{twoparts}). Therefore by combining the curves in Figs.
\ref{fig:3} and \ref{fig:5} we come to the pattern presented in
Fig. \ref{fig:6}, which is the detailing of Fig. \ref{fig:1}. Each
potential curve drawn for a certain value of the magnetic field
approaches, as the distance from the charge along the $x_3$-axis
grows, the corresponding (dashed) curve transferred from Fig.
\ref{fig:5}. But even prior to this, the latter approaches the
thick dashed curve, which is the common envelope of the curves in
Fig. \ref{fig:5} and the whole potential curves in Figs.
\ref{fig:1} and \ref{fig:6}.
\begin{figure}[htb]
  \begin{center}
   \includegraphics[bb = 0 0 405 210,
    scale=1]{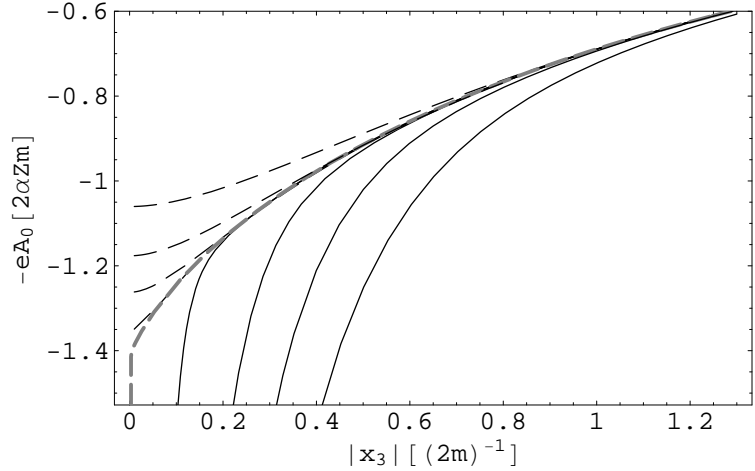} \caption{
    %Detailing Fig. \ref{fig:1}.
    Four electron energy $-eA_0(x_3,0)$ curves in
     the modified potential (\ref{bessel}) for $b=10^6,\,
     10^5,\,3\times 10^4,\,{\rm and}\,
10^4$ (thin solid lines from left to right) approaching their
corresponding long-range parts $-eA_{\rm l.r.}(x_3,0)$, Eq.
(\ref{long}), shown in Fig. \ref{fig:5} (four dashed lines from
bottom to top). Thick dashed broken line corresponds to  the string
potential $A_0(x_3,0)\left.\right|_{b=\infty}$. Its vertical
fragment symbolizes the $\delta$-function (\ref{deltalimit}).}
\label{fig:6}
  \end{center}
\end{figure}

 We continue by studying the long-range part of the potential along the
 string, Eq. (\ref{envelope}).
 To separate the
part independent of the fine-structure constant $\alpha$ the
internal integral here is integrated by parts to yield:
\bee\label{parts}\int_0^\infty\left({\rmd\over\rmd
\widetilde{k}_\perp^2}\ln
\frac{\widetilde{k}_\perp^2+{2\alpha\over\pi}\exp\left({-\widetilde{k}_\perp^2\over
2}\right)T(y)}{\widetilde{k}_\perp^2+{2\alpha\over\pi}\exp\left({-\widetilde{k}_\perp^2\over
2}\right)}\right)\frac{\;\rmd
\widetilde{k}_\perp^2}{1+{\widetilde{k}_\perp^2\over 2}
}\hspace{4.5cm}\nonumber\\= -\ln T\left(y\right)+{1\over
2}\int_0^\infty\frac{\rmd
\widetilde{k}_\perp^2}{\left(1+{\widetilde{k}_\perp^2\over
2}\right)^2}\ln
\frac{\widetilde{k}_\perp^2+{2\alpha\over\pi}\exp\left({-\widetilde{k}_\perp^2\over
2}\right)T(y)}{\widetilde{k}_\perp^2+{2\alpha\over\pi}\exp\left({-\widetilde{k}_\perp^2\over
2}\right)}.\eend Then (\ref{envelope}) becomes
\bee\label{envelope2}\left.A_{\rm l.r.}(x_3,0
)\right|_{b=\infty}=\left.A_{\rm
l.r.}(x_3,0)\right|_{b=\infty,\alpha=0}\hspace{7cm}\nonumber\\+\frac{q}{2(2\pi)^2}\int_0^\infty
\int_0^\infty\frac{\cos(k_3x_3)\rmd k_3\rmd
\widetilde{k}_\perp^2}{\left(1+{\widetilde{k}_\perp^2\over
2}\right)^2}\ln\left\{1+
\frac{{2\alpha\over\pi}\exp\left({-\widetilde{k}_\perp^2\over
2}\right)\left[T\left({k_3^2\over
4m^2}\right)-1\right]}{\widetilde{k}_\perp^2+{2\alpha\over\pi}\exp\left({-\widetilde{k}_\perp^2\over
2}\right)}\right\},\hspace{1cm} \eend where the first term (it is
worth recalling here that within the integration limits $T(y)$ is a
positive function, lesser than unity) \bee\label{alpha0}\left.A_{\rm
l.r.}(x_3,0)\right|_{b=\infty,\alpha=0}= -\frac {q}{(2\pi
)^2}\int_0^{\infty}\cos (k_3x_3)\ln \left[T\left({k_3^2\over
4m^2}\right)\right]\rmd k_3\eend is independent of $\alpha$, whereas
the second term behaves as \bee\label{smallalpha}\frac
{q}{4\pi^3}(-\alpha\ln\alpha)\int_0^{\infty}\cos (k_3x_3)
\left[T\left({k_3^2\over 4m^2}\right)-1\right]\rmd k_3.\eend  when
$\alpha$ tends to zero, i.e. is nonanalytic in $\alpha=0$. The
reason for the nonanalyticity and for the nonvanishing of
$\left.A_{\rm l.r.}(x_3,0)\right|_{b=\infty,\alpha=0}$ is in that
that a chain of diagrams has been as a matter of fact summed when
solving the Dyson-Schwinger equation that led to the expression for
the photon Green function (\ref{decomposition}) with the one-loop
polarization operator  (\ref{2}) substituted into it. In the result
(\ref{A0}) thus obtained the two limits $b=\infty$ and $\alpha=0$
are not permutable.

Eq. (\ref{alpha0}) may be referred to as  a fitting approximation
for the envelope (\ref{envelope}),  simpler than
(\ref{envelope2}). It is presumably useful for making rough
estimates with the accuracy to $(-\alpha/\pi)\ln\alpha =0.011$. It
might have been obtained if the exponential
$\exp(-\widetilde{k}_\perp^2/2)$ in (\ref{envelope}) had been
merely replaced by unity. The integral (\ref{alpha0}) is
converging at the both integration limits due to the asymptotic
properties of the function $T(y)$ indicated below its definition
(\ref{T}) and represents a function that decreases at large
longitudinal distances following the Coulomb law $(q/4\pi x_3).$

The limiting curve (\ref{envelope}), or (\ref{envelope2}),
(\ref{alpha0}) for the long-range part of the potential
(\ref{long}) crosses the axis $x_3=0$ in the point $\left.A_{\rm
l.r.}(0,0
)\right|_{b=\infty}=(1.4240-0.0088)(qm/2\pi)=1.4152(qm/2\pi)$. %The
%value $\alpha=1/137.05$ was used in computing the integral
%(\ref{envelope2}) at $x_3=0$ ({\em and integration over
%d$\widetilde{k}_\perp^2$ was performed first.  If the order of
%integrations is reversed, one gets the figure 0.008709 in place of
%0.00884}).
Here two numerical contributions from the first and the second
terms in (\ref{envelope2}) are presented separately. It is
intriguing how close the numerical coefficient in the expression
for the intercept of the envelope and the $x_3=0$ axis is to
$\sqrt{2}=1.4142$. More precise value of $\sqrt{2}$ would be
achieved by the infinite-magnetic-field limit of the long-range
part of the potential in the point where its charge is located, if
the fine-structure constant were 1/121. Higher-loop calculations
may improve this figure.

Identifying the above-calculated $\alpha$-dependent coefficient
1.4152  supposedly  with $\sqrt{2}$, an interesting observation
would follow: if the charge $q$ is taken equal to the electron
charge $e$, $(Z=1)$, the string potential undergoes the increment
between the point $x_3$ where the charge is located and the
infinitely remote point $x_3=\infty$, equal to
\bee\label{potdif}\Delta A_{\rm l.r.}(x_3)\equiv A_{\rm
l.r.}(0,0)\left.\right|_{b=\infty}-A_{\rm
l.r.}(\infty,0)\left.\right|_{b=\infty}=
  \frac e{\pi\sqrt{2}\; \lambda_{\rm C}}, \eend where $\lambda_{\rm C}=m^{-1}$ is the Compton
  length. This "work function over a unit charge" differs from the
  photon mass (\ref{mass2}) in that the Larmour dimensioning has
  been replaced by the Compton one.

In the interval $|x_3|\ll (2m)^{-1}$ the envelope curve
(\ref{envelope}) looks roughly in Fig. \ref{fig:1} as a linearly
growing potential, the same as this is believed to be the case for
the confining potential along quark-antiquark string in QCD in the
limit of zero lattice spacing. We may say that in QED the
"confinement" occurs within distances smaller than the Compton
length, whereas for larger distances - thanks to the fact that the
infrared behavior in QED is weaker than in QCD - the growth of the
potential ceases and it approaches the zero value along the Coulomb
asymptote.  As a matter of fact the growth of the potential is only
nearly linear.

To establish its true character consider the difference
$\left.A_{\rm l.r.}(x_3,0 )\right|_{b=\infty}-\left.A_{\rm l.r.}(0,0
)\right|_{b=\infty}$ and change to the new integration variable
$u=k_3x_3$ in the integrals (\ref{alpha0}) and (\ref{envelope2}).
Then the argument of the function $T(y)$ becomes $(u^2/4m^2x_3^2)$
and should be considered as large, once $4m^2x_3^2\ll 1$. According
to Eq. (\ref{T}) for large $y$ one has $[T(y)-1]\simeq (-\ln
2y/2y)$. As long as this tends to zero with $y\rightarrow\infty$, we
may substitute $\ln (1+x)\approx x,\; x\ll 1$ for the logarithms in
(\ref{alpha0}) and (\ref{envelope2}). In this way we
obtain\bee\label{confinement}\left.A_{\rm l.r.}(x_3,0
)\right|_{b=\infty}-\left.A_{\rm l.r.}(0,0
)\right|_{b=\infty}=\frac{qm}{4\pi}\left(1-{\alpha\over\pi}
f(\alpha)\right)2m|x_3|\left[\ln (2m|x_3|)-{1\over 2}\ln
2+\gamma-1\right],\quad \eend where
\bee\label{f(alpha)}f(\alpha)=\int_0^\infty\frac{\exp\left({-\widetilde{k}_\perp^2\over
2}\right)\rmd
\widetilde{k}_\perp^2}{\left(1+{\widetilde{k}_\perp^2\over
2}\right)^2\left[\widetilde{k}_\perp^2+{2\alpha\over\pi}\exp\left({-\widetilde{k}_\perp^2
\over 2}\right)\right]},\quad f\left({1\over
137.036}\right)=4.533,\quad
\left.f(\alpha)\right|_{\alpha\rightarrow 0}\simeq -\ln\alpha,
\qquad \eend and $\gamma=0.577$ is the Euler constant. We have made
use of the two standard integrals
\cite{ryzhik}\bee\label{standard}\int_0^\infty\frac{(\cos u
-1)}{u^2}~\rmd u=-{\pi\over 2},\qquad\qquad\int_0^\infty\frac{(\cos
u -1)}{u^2}\;\ln u\;\rmd u={\pi\over 2}(\gamma -1).\eend Finally,
for small distances along the string, $2|x_3|m\ll 1$, the long-range
potential has the form:\bee\label{confinement2}\left.A_{\rm
l.r.}(x_3,0 )\right|_{b=\infty}=\frac{qm}{2\pi}[1.4152+0.495\times
2m|x_3|(\ln (2m|x_3|)-0.77)], \eend This should be additively
combined with the $\delta$-function (\ref{deltax3}), to which the
short-range part $A_{s.r.}(x_3,0)$ is reduced in the limit
$b=\infty$, to form the string potential. It is this potential that
is responsible for forming the spectrum of an atom in the strong
magnetic field, to consideration of which we are proceeding.

\section{Radiative shift of electron ground-state energy in a
hydrogenlike atom in a strong magnetic field} In this section we
shall study how the ground-state energy of a hydrogenlike atom at
rest in a strong magnetic field is modified by the radiative
corrections to the Coulomb potential considered above.

The nonrelativistic electron in an atom, whose nucleus has the
charge $q=Ze$, is described by the one-dimensional Schr$\ddot{\rm
o}$dinger equation \cite{elliott}\bee\label{sch}\hspace{-2cm}
-\frac 1{2m}\frac{\rmd^2\Psi (x_3)}{\rmd x_3^2}-eA_0(x_3,x_\perp
=0)\Psi(x_3)=E\Psi(x_3), \qquad |x_3|>L_{\rm B}=(eB)^{-1/2},\eend
if the atom does not move transverse to the magnetic field - which
is the case as long as we are interested in its ground state. The
one-dimensional \sch  (\ref{sch}) is valid in the region
$|x_3|>L_{\rm B}$ and is efficient provided that $L_{\rm B}\ll
a_{\rm B}/Z$, where $a_{\rm B}=1/m\alpha$ is the Bohr radius.

If the unmodified Coulomb potential (\ref{free}) taken at
$x_\perp=0$ is used for $A_0(x_3,0)$ in equation (\ref{sch})
($q=Ze$ henceforward), the
ground-state energy value  \bee\label{LE} E_0 %=-\frac {2Z^2}{ma_{\rm
%B}^2}\ln^2\frac{2L_{\rm B}}{a_{\rm B}}
=-2Z^2\alpha^2m\ln^2\frac{\sqrt{b}}{2\alpha Z},\eend that follows
\cite{elliott,haines} from equation (\ref{sch}) is unbounded from
below, in other words tends to negative infinity as the magnetic
field grows. The reason is that the one-dimensional Coulomb
potential is too singular, the singularity being regularized by
the Larmour radius. In equation (\ref{sch}) this regularization
acts as the cut-off of the definition region $|x_3|>L_{\rm B}$ of
equation (\ref{sch}). The regularization is lifted by letting $b$
tend to infinity, $L_{\rm B}\to 0,$  and hence $E_0\to -\infty$.
On the contrary, the radiation corrections studied here, yielded
the conclusion that the Coulomb $q/(4\pi x_3)$ singularity of the
one-dimensional potential in $x_3=0$ had been changed to the
$\delta$-function (\ref{deltax3}). This sort of singularity is not
expected to cause an unboundedness of the energy spectrum. To
confirm this, we solve in Appendix II the \sch (\ref{sch}) with a
 potential that models the short-range part $A_{\rm s.r.}(x_3,0)$
 (\ref{short}) alone and also tends to $\delta$-function as
 $b\to\infty$. The resulting ground-state energy approaches  in this limit
 the finite, magnetic-field-independent value given by Eq. (\ref{energy2})
 of Appendix II. %:\bee\label{energy2} -
%2Z^2\alpha^2m\ln^2\sqrt{\frac\pi{2\alpha}}=-7.686\times
%10^{-4}mZ^2=-389.3~{\rm eV}\times Z^2.\eend
The genuine level must be significantly lower due to impact of the
long-range part of the potential $A_{\rm l.r.}(x_3,0)$
(\ref{long}) shown in Fig. \ref{fig:5}.

\subsection{Extremely large magnetic fields}

To estimate the ground-state energy $E$ in the limiting case
$b\rightarrow\infty$ we apply here the shallow-well approximation of
Ref. \cite{QM}, appropriate since the electron potential $V=-e
A_0(x_3,x_\perp=0)$ has a small depth ($|V|\ll (ma^2)^{-1}$, where
$a$ is the range of the forces in the well). In this case, the value
of $E$ may be estimated as \bee\label{Eshallow} E\simeq
-2m\left(\int^\infty_0 eA_0(x_3,0) dx_3\right)^2.\eend Here it is
taken into account that the electron potential is symmetrical,
$A_0(-x_3,0)=A_0(x_3,0)$.

To find first the contribution of the long-range part into
(\ref{Eshallow}) rewrite (\ref{long}) as ($q=eZ$,
$\widetilde{k}_\perp m\sqrt{b}=k_\perp,\; 2mk=k_3$
)\bee\label{long2}eA_{\rm l.r.}(x_3,0)=2\alpha Zm\frac 1{\pi}
\int_0^\infty \int_0^\infty \cos (2mkx_3)\,\rmd
k\nonumber\hspace{5cm}\\\times \left[\frac{1}{
\widetilde{k}_\perp^2+\frac 4{b}k^2+\frac{2\alpha}{\pi }\exp \left(-
\frac{\widetilde{k}_\perp^2}{2}
\right)T\left(k^2\right)}-\frac{1}{\widetilde{k}_\perp^2+\frac 4{b}
k^2+\frac{2\alpha }{\pi}\exp\left(- \frac{\widetilde{k}_\perp^2}{2}
\right)}\right]\rmd \widetilde{k}_\perp^2.\eend The $b=\infty$ limit
of this expression is Eq. (\ref{envelope}) or (\ref{envelope2}). In
the problem under consideration the potential falls following the
Coulomb law, and hence, according to \cite{QM}, the upper
integration limit in (\ref{Eshallow}) should be replaced by the Bohr
radius $a_{\rm B}=(m\alpha)^{-1}$. Then the contribution of the
long-rang part (\ref{long2}) into the ground-state energy is
determined by the integral\bee\label{contrib1} \int_0^\frac
1{m\alpha} eA_{\rm l.r.}(x_3,0)\rmd x_3=\alpha Z\frac
1{\pi}\int_0^\infty\int_0^\infty\sin
\left(\frac{2k}\alpha\right)\frac 1{k} \left[\frac{1}{
\widetilde{k}_\perp^2+\frac 4{b}k^2+\frac{2\alpha}{\pi }\exp \left(-
\frac{\widetilde{k}_\perp^2}{2}
\right)T\left({k^2}\right)}\right.\nonumber\\-\left.\frac{1}{\widetilde{k}_\perp^2+\frac
4{b} k^2+\frac{2\alpha }{\pi}\exp\left(-
\frac{\widetilde{k}_\perp^2}{2} \right)}\right]\rmd k\,\rmd
\widetilde{k}_\perp^2. \hspace{3cm}\eend Analogously, the
contribution of the short-range part (\ref{short}) into the ground
state energy according to (\ref{area2}) is the
magnetic-field-independent constant
\bee\label{area4}\int^\infty_{L_{\rm B}} eA_{\rm s.r.}(x_3,0)\rmd
x_3=\alpha Z~2.18\,.\eend
%The ground-state energy (\ref{Eshallow})
%is plotted by a computer using Eq. (\ref{contrib1}), (\ref{area4})
%and (\ref{twoparts}) in Fig. \ref{fig:8} against the magnetic
%field. Its constant limiting value can be found by integrating Eq.
%(\ref{envelope2}).

As a matter of fact, only the contribution of the first,
$\alpha$-independent term
(\ref{alpha0})\bee\label{contrib2}\int_0^{\frac 1{m\alpha}} eA_{\rm
l.r.}(x_3,0)\left.\right|_{b=\infty,\;\alpha=0}\rmd x_3=-\alpha
Z\frac 1{\pi}\int_0^\infty\sin\left(\frac{2k}\alpha\right)\ln
T(k^2)\frac{\rmd k}k=\alpha Z~6.392 \eend is important, whereas the
second term in (\ref{envelope2}) only corrects the value 6.392 in
the third decimal number (+$6\cdot 10^{-3}$). Combining
(\ref{contrib2}) with (\ref{area2}) we get from (\ref{Eshallow}) the
finite value of the energy level of a hydrogenlike atom in the limit
$b=\infty$ \bee\label{limenergy} E_{\rm
lim}=-2mZ^2\alpha^2~73.6=-Z^2\times 4.0 ~{\rm keV}.\eend This result
reproduces with good accuracy the value obtained by us earlier
\cite{PRL07} with the use of a graphically fitted formula in place
of Eq. (\ref{envelope2}).

The Loudon-Elliott energy (\ref{LE}) would overrun the limiting
energy (\ref{limenergy}) already for the magnetic field as large as
$b=6600$, when yet the proton size $R\sim 10^{-13}$cm remains much
smaller than the Larmour length, $R\ll L_{\rm B}$. The ground level
reaches 92\% of its limiting value for $b=5\times 10^4$. After the
magnetic field reaches the value $b=1.5\times 10^5$, when $R$ and
$L_{\rm B}$ equalize, the Coulomb potential is cut off at the proton
size, $x_3=R$. Setting $L_{\rm B}=R$ in (\ref{LE}) we would get the
minimum value for the Loudon-Elliott energy ($Z=1$) to be $-5.6$
keV, which is essentially lower than (\ref{limenergy}).

\subsection{Moderate magnetic fields}
For moderate magnetic fields lying in the range $1\ll b \ll 10^3$
the additive radiative correction to the
 Coulomb law, as calculated in Subsection B of Appendix I keeping
the first power of $\alpha b/2\pi$ in the power series expansion of
$A_0(x_3,0)$, \bee\label{uehling3}\Delta A_0(x_3,0)\simeq
\frac{q\alpha b m}{8\pi^2}\int_0^{\pi/2} \exp \left(
-\frac{2m|x_3|}{\cos\phi}\right)\cos^2\phi~\rmd\phi\eend may be
considered as perturbation. Therefore, the radiative shift to the
ground-state energy level can in this case be calculated using the
purely Coulomb (normalized) wave function
\cite{haines}\bee\label{wf}\Psi(x_3)=\frac{1}{\sqrt{\delta_0 a_{\rm
B}}} \exp \left({-\frac{|x_3|}{\delta_0 a_{\rm B}}}\right)\eend as
unperturbed. Here $\delta_0$ is the "quantum defect" for the Coulomb
 problem \bee\label{delta_0}
\delta_0=\frac 1{Z}\ln\left(\frac b{4\alpha^2Z^2}\right)
%E_0=-\frac{m\alpha^2_0}{2\delta^2}.
\eend Calculating the average of (\ref{uehling3}) multiplied by $e$
with the wave function (\ref{wf}) we find the perturbation to the
Loudon-Elliott ground-state energy ($q=Ze$
)\bee\label{perturbed}E-E_0=\frac{Z\alpha^2 b m}{2\pi a_{\rm
B}\delta_0}\int_0^{\frac \pi{2}}\frac{\cos^3\phi~\rmd\phi}{m+\frac
{\cos \phi}{a_{\rm B}\delta_0}}=\frac{Z\alpha^2 b m}{2\pi
}\int_0^{\frac
\pi{2}}\frac{\cos^3\phi~\rmd\phi}{\cos\phi+\delta_0/\alpha}.\eend
One sees that for magnetic fields within the scope of applicability
of the expansion in powers of $\alpha b,$ $1\ll b\ll 10^3$, where
(\ref{uehling3}) is valid, the quantum defect  $\delta_0\gg \alpha$,
or $(1/Z\alpha)-8.454\gg \ln (b/Z^2),$ provided that $Z\leq 11$.
Then, for the light hydrogenlike atoms Eq. (\ref{perturbed}) can be
further simplified to
\bee\label{pertutbed2}\hspace{-1.5cm}E-E_0=%\frac{Z^2\alpha^2 b
%m}{3\pi }\frac {\alpha}{Z\delta}=
\frac{Z^2\alpha^3 b m}{3\pi
}\ln\frac
b{4\alpha^2Z^2}=Z^2b~0.18~\large(\frac{1}{8.454}\ln\frac{b}{Z^2}
+1\large)~{\rm eV}.\eend

\subsection{Relativistic corrections to electron motion}
%%%%%%%%%%%%%%%%%%%%%%%%%%%%%%%%%%%
%%%%%%%%%%%%%%%%%%%%%%%%%
%%%%%%%%%%%%%%%
The value (\ref{limenergy}) makes about $1\%$ and more of the
electron rest mass,
 hence the question about relativistic corrections may arise.

When the Dirac equation with stationary Coulomb potential is considered
in infinitely growing magnetic field,
the effect of unlimited lowering of the energy level down to $-\infty$
is enhanced as compared to the Schr$\ddot{\rm o}$dinger equation due to the
known fact \cite{blp} that the potential is squared after the Dirac equation is reduced
to one-component second-order differential equation. Therefore, we should face
an one-dimensional second-order equation with the stronger singularity  $(\alpha Z/x_3)^2$,
 apart from
the singularity  $\alpha Z/x_3$ already present in (\ref{sch}). For
this reason one may expect that the ground-state energy would tend
to negative infinity faster than the logarithm squared in
(\ref{LE}). Anyway, according to the (numerical part of the)
analysis in Ref. \cite{semikoz}, it rather sharply approaches the
border of the lower continuum $E=-m$,
 where the instability with respect to free positron production opens, analogous to what happens
(without any magnetic field)
for nuclei with the supercritical charge $Z>137$ (to be more precise, $Z>170$, once the finite
size of the nucleus is taken into account) \cite{supercritical}. Whereas for infinite magnetic
field the unlimited sinking of the level occurs
already for infinitesimal Coulomb attraction $Z\alpha\rightarrow 0$, for large, but finite magnetic field
the Coulomb-induced quadratic singularity is cut off at the Larmour length, hence the ground
level reaches the lower continuum at finite $Z\alpha$. The dependence of the corresponding
 critical value of $Z$ on the magnetic field was found long ago by Oraevskii, Rez and Semikoz
 \cite{semikoz}, who claimed, for instance, that  already for the values of magnetic fields
$b=10^2$ to $10^3$ (that may exist near neutron stars according to
the estimates available at present time, see below), the critical
value of the nuclear charge lies within the reasonable range in the
Periodic Table $Z= 55$ to $Z=90$. This result is to be reconsidered
now that we have established the important alteration in the
singular behavior of the modified Coulomb potential proved to be
crucial for  the Schr$\ddot{\rm o}$dinger equation with huge
magnetic field.

The next level of relativistic description of atomic (or
positronium) spectrum based on the static potential would be that
via the Bethe-Salpeter equation with the so-called equal-time Anzatz
wherein the recoil of the point source of the electrostatic field (a
nucleus or a positron) is taken into account, but the retardation
effects in the relative motion of the electron and the nucleus
(positron) are disregarded. The corresponding results established in
\cite{leinson,ShUs,lai} should be also subjected to revision. (This
statement does not concern the conclusions about  the effect of
photon capture through positronium formation in the pulsar
magnetospheres made in \cite{ShUs}).

The matters stand differently when very deep relativistic effects
are dealt with. The latter come into play for magnetic fields tens
of orders of magnitude higher than those for which the asymptotic
limit in the present context is saturated (i.e., than, say,
$b=10^{10}$). Retardation effects make the static potential an
insufficient quantity to take on the responsibility for forming
bound states,
 since the full electromagnetic interaction is mediated
 by all the three photon modes in (\ref{decomposition}). Unlike
(\ref{2}), the polarization operator
eigenvalues $\kappa_{1,3}$ of two other modes do not include
\cite{melrose1,kratkie,shabtrudy,skobelev} the fast-growing factor
$b$, and hence the interaction singular on the light cone
$x_0^2-{\bf x}^2=0$ characteristic of the free photon propagator is
not  suppressed in these modes. Correspondingly, the infinite
deepening of the
 energy level, considered in our papers \cite{prl} for positronium atom in a magnetic field using
the Bethe-Salpeter equation without the equal-time Ansatz, survives
the radiative corrections, as well as the effect of vacuum instability that occurs at the magnetic
 field value about $b=1.6\times 10^{28}$. This indicates the existence of a maximum magnetic
field in quantum electrodynamics. Note that contrary to the Dirac
case \cite{semikoz}, where the critical magnetic field is determined
by the large factor $\exp(1/\alpha Z)$, for the Bethe-Salpeter case
($Z=1$ for positronium) we got the factor $\exp(1/\alpha^{1/2}) $.

From Eq. (\ref{pertutbed2}) the relative correction to the ground
state energy for moderate magnetic fields is
\bee\label{relative}\frac{E-E_0}{|E_0|}= \frac{\alpha
b}{3\pi}\left(\ln\frac{\sqrt{b}}{2\alpha Z}\right)^{-1}.\eend For
$Z=1$ this correction, when extrapolated (though unrighteously) down
to the value $b=0.27$, is of the same order of magnitude $(5.8\times
10^{-5})$ as the relativistic relative correction $(2.8\times
10^{-5})$ calculated by Goldman and Chen \cite{goldman} basing on
the Dirac equation for this - largest in their analysis - value of
$b$. The same situation retains, if the results of these authors are
linearly extrapolated (using their two largest values of $b$ for
$Z=1$) into the region of larger $b$, $ 1\ll b\ll 10^3$, wherein
(\ref{relative}) is valid. Therefore, already for
 magnetic fields far from critical fields causing the free positron
production instability the impact of vacuum polarization is
at least no less important than relativism introduced by the use of the Dirac, instead of
the Schr$\ddot{\rm o}$dinger equation.
%%%%%%%%%%%%%%%%%%%%%%%%%%%%%%%%%%%%%%%
 %%%%%%%%%%%%%%%%%%%%%%%%%%%%%%%%%%%%%%%%%%%%%%%%%%%%%
\section{Discussion}
In this paper we have shown that the electric field of a pointlike
charge placed in a strong magnetic field ($b=B/B_0\gg 1$) may be
significantly modified by the vacuum polarization, especially if
$b\gtrsim 3\pi \alpha^{-1}\simeq 10^3$. At present, it is commonly
accepted that many compact astronomical objects identified with
neutron stars are strongly magnetized. For soft gamma-ray repeaters
and anomalous X-ray pulsars, for instance, the strength of the
surface magnetic field is estimated as $\sim 10^{14}-10^{15}$~G
\cite{TD95}. Several radio pulsars with similar surface magnetic
fields have been recently discovered \cite{kaspi}. More strong
magnetic fields ($B\sim 10^{16}-10^{17}$~G or even higher) are
predicted to exist at the surface of cosmological gamma-ray bursters
if they are rotation-powered neutron stars similar to radio pulsars
\cite{U92}. The modification of the Coulomb law should affect the
electric fields of an atomic nuclei and electrons placed in such a
strong magnetic field. The electric field of a particle is one of
its fundamental features. Therefore, at the surface of neutron stars
with extremely strong magnetic fields many properties of matter
(including individual atoms and molecules) and various physical
processes (such as radiation of particles) where the electric field
of particles is important (for a review on physics of strongly
magnetized neutron stars, see \cite{HD06}) may be changed
substantially by the present modification of the Coulomb law. One of
such changes is discussed in Section~V where we have come to
negation of
%studied the ground-state energy of a hydrogenlike atom by
%considering the \sch with the vacuum-polarization-modified
%potential. Unless the vacuum polarization is taken into account,
the standard result \cite{elliott}, referred to in many speculations
on behavior of matter on the surface of strongly magnetized neutron
stars (e.g., \cite{HD06} and references therein), that the
ground-state energy tends to negative infinity
as the magnetic field unlimitedly grows. % The conclusion of the
%present paper is that if the modified potential is used in the
%\sch the ground-state energy remains finite simply because the
%singularity of the modified potential in the
%infinite-magnetic-field limit ($b=\infty$) at the origin, $|{\bf
%x}|=0$, has the $\delta$-function character.

%The fact that in the $b=\infty$ limit the
%vacuum-polarization-modified potential is concentrated on the axis
%$x_\perp =0$, making an infinitely thin tube or string, may be of
%great interest as the first evidence of dimensional reduction of
%the photon sector of quantum electrodynamics.
We hope that the modification of the Coulomb potential described in
the present paper %this fundamental effect may also
 may lead to observational appearances in neutron stars with
extremely strong magnetic fields.
%The appearance of an one-dimensional string  in the
%infinite-magnetic-field limit may be of interest as the first
%evidence of dimensional reduction of the photon sector of quantum
%electrodynamics.
As for the results relating to much larger magnetic fields,
infinite in the limit, such as the QED string formation, these may
be of fundamental importance as introducing a nonempty
magnetic-field-independent two-dimensional theory in virtue of
dynamical dimensional reduction from 4-dimensional quantum
electrodynamics.

\begin{acknowledgments} This work was
supported by the Russian Foundation for Basic Research (project no
05-02-17217) and the President of Russia Programme
(LSS-4401.2006.2), as well as by the Israel Science Foundation of
the Israel Academy of Sciences and Humanities.
\end{acknowledgments}
\section*{Appendix I}
\subsection{Asymptotic expansion around the singular point ${\bf x}=0$}
To consider the behavior of the potential  near its pointlike source
let us add to and subtract from (\ref{bessel}) the standard Coulomb
potential (\ref{free2}) in the form \bee\label{shortperp}A_0^{\rm
C}({\bf x})=\frac q{(2\pi)^3}\int\frac{\rme^{-\rmi{\bf
kx}}\rmd^3k}{{\bf k}^2} \nonumber\\ =\frac
{q}{2(2\pi)^2}\int_0^\infty J_0(k_\perp
x_\perp)\left(\int_{-\infty}^\infty\frac{\rme^{-\rmi k_3x_3}\rmd
k_3}{ k_\perp^2+k_3^2}\right)\rmd k_\perp^2=\frac 1{4\pi}\frac
q{\sqrt{x_\perp^2+x_3^2}}\eend so that \bee\label{difference1}
A_0({\bf x})=A_0^{\rm C}({\bf x})-\Delta A_0({\bf x}), \eend where
\bee\label{difference2} \Delta A_0({\bf x})=
\frac{q}{2(2\pi)^2}\int_0^\infty J_0(k_\perp
x_\perp)\int_{-\infty}^\infty\left(\frac{\rme^{-\rmi k_3x_3}}{
k_\perp^2+k_3^2}-\frac{\rme^{-\rmi k_3x_3}}{
k_\perp^2+k_3^2-\kappa_2(0,k_3^2,k_\perp^2)}\right)\rmd k_3\rmd
k_\perp^2.\eend  Note that the function $\Delta A_0(x_3,x_\perp)$ is
an entire function of $x_\perp$, since the exponential in (\ref{2})
provides convergence of the integral (\ref{difference2})
for any complex value of this variable. %Once the correction
%(\ref{difference2}) to the Coulomb potential, unlike
%(\ref{bessel}), does not diverge in the point $x_3=x_\perp=0$, it
%is not a function singular in the origin. This will allow later
%formal manipulations, like limiting transitions in the integrand,
%with it, inadmissible as long as (\ref{bessel} is concerned.
Keeping quadratic terms in the power series expansion of
$J_0(k_\perp x_\perp)$ and $\exp (-\rmi k_3x_3)$ in
(\ref{difference2}) we obtain  the first three terms  of the
asymptotic expansion of the potential (\ref{bessel}) near the
origin $x_3=x_\perp=0$ \bee\label{expand}A_0({\bf x})\sim
\frac{q}{4\pi}\left(\frac 1 {|\bf
x|}-2m(C-(2mx_\perp)^2C_\perp-(2mx_3)^2C_\parallel)\right), \eend
where $C$, $C_\perp$ and $C_\parallel$ are  dimensionless positive
constants depending on the external field:\bee\label{constant}
C\equiv \frac{2\pi}{qm}\Delta A_0(0) =\frac{\alpha b
m}{\pi^2}\int_0^\infty
T\left(\frac{k_3^2}{4m^2}\right)\int_0^\infty \frac
{\exp\left(-\frac{k_\perp^2}{2m^2b}\right) \rmd k_\perp^2}
{(k_\perp^2+k_3^2)(k_\perp^2+k_3^2-\kappa_2(0,k_3^2,k_\perp^2))}
\rmd k_3 ,\eend \bee\label{constantperp}\hspace{-1.5cm}C_\perp
=\frac{\alpha b }{16m\pi^2}\int_0^\infty
T\left(\frac{k_3^2}{4m^2}\right)\int_0^\infty \frac
{k_\perp^2\exp\left(-\frac{k_\perp^2}{2m^2b}\right) \rmd
k_\perp^2}
{(k_\perp^2+k_3^2)(k_\perp^2+k_3^2-\kappa_2(0,k_3^2,k_\perp^2))}
\rmd k_3 ,\eend  \bee\label{constantpar}\hspace{-1.5cm}C_\parallel
=\frac{\alpha b }{8m\pi^2}\int_0^\infty
T\left(\frac{k_3^2}{4m^2}\right)\int_0^\infty \frac
{k_3^2\exp\left(-\frac{k_\perp^2}{2m^2b}\right) \rmd k_\perp^2}
{(k_\perp^2+k_3^2)(k_\perp^2+k_3^2-\kappa_2(0,k_3^2,k_\perp^2))}
\rmd k_3 ,\eend Thanks to the exponential factor the integrals
over $k_\perp^2$ here are fast converging. The resulting functions
decrease for large $k_3$ as $\sim 1/k^4_3$, so the remaining
integrals over $k_3$ in (\ref{constant}), (\ref{constantperp}),
(\ref{constantpar}) converge, bearing in mind that $T$ is a
bounded function. The inequality $C_\perp\neq C_\|$ implies the
anisotropy.

The values of the coefficients %$C, C_\perp, C_\parallel$
%calculated following
(\ref{constant}), (\ref{constantperp}), (\ref{constantpar})
calculated for four values of the magnetic field $b=10^4,$
$b=10^5$, $b=10^6$ and $b=10^{10}$ are listed in the Table
\vspace{1cm}
 \hspace{2cm}\begin{tabular}{ccccc} $b$
&10$^4$&$10^5$&$10^6$&$10^{10}$\\$C$ &2.21&9.08&31.37&32.70$\times
10^2$\\$C_\perp$&75.9&2.58$\times
10^3$&8.38$\times 10^4$&8.49$\times 10^{10}$\\
 $C_\parallel$&174.3&5.55$\times 10^3$&1.76$\times 10^5$&1.67$\times 10^{11}$
\end{tabular}\\

\vspace{1cm}

 To find the asymptotic behavior of the constant $C$
(\ref{constant}) as $b\to\infty$, we may use first the
representation (\ref{universal}) for the short-range part of the
potential. The corresponding contribution $C_{\rm s.r.}$ into $C$ is
\bee\label{constant1} C_{\rm s.r.}=\sqrt{\frac{\alpha b}{2\pi}}
\int_0^\infty\left( 1-\frac {u}{\sqrt{u^2+\exp (-\frac
\alpha{\pi}u^2)}}\right)\rmd u.\eend By restricting the upper
integration limit to the value $\sqrt{\pi/\alpha}$ and substituting
unity for the exponential the integral in (\ref{constant1}) can be
estimated as approximately equal to $(1-\sqrt{\alpha/2\pi})=0.996$.
A computer calculation results in the value 0.9595. Correspondingly
\bee\label{C0}C_{\rm s.r.}\simeq 0.9595\sqrt{\frac{\alpha
b}{2\pi}}.\eend  The resulting values of $C_{\rm s.r.}$ for $b=10^4,
10^5, 10^6, 10^{10}$ are 3.27, 10.34, 32.70, 32.70$\times 10^2,$
correspondingly, to be compared with the exact values given in Table
above. The coincidence is the better the larger the field. It
improves if the (negative) contribution to $C$ of the long-range
part of the potential is added to to this row of  numbers. (The
absolute value of) the latter is a decreasing function of $b$ that
takes the limiting value $C_{\rm l.r.}=-(2\pi/qm)A_{\rm
l.r.}(0,0)\left.\right|_{b=\infty}=-1.4152$ according to Subsection
B of Section IV. Note that if just the Yukawa law (\ref{yuk}) is
accepted for the potential we would deduce that for strong fields
$C\simeq \frac M{2m}=\sqrt{\frac{\alpha b}{2\pi}}$ asymptotically.
For the four values of  the external field
$b=10^4,\;10^5,\;10^6,\;10^{10}$ the values of $C$ calculated
following the Yukawa law are: 3.41, 10.78, 34.01 and $34.08\times
10^2$.
\subsection{Modified Coulomb potential for less huge magnetic fields}
 Consider the "moderate" values of the magnetic field in the interval
$10^3\gg b\gg 1$, so that although $b$ is large, but $(\alpha
b/2\pi)=1.16\times 10^{-3} b$ is still much less than unity. We
shall present here the vacuum polarization correction to the
Coulomb potential, which in this case is small.

 One may neglect
$\kappa_2$ in the denominator of (\ref{difference2}) after the
difference in it is completed and we obtain a magnetized vacuum
analog of the Uehling-Serber potential \cite{blp}. Contrary to the
Uehling-Serber potential that is of the order of $\alpha$,
 its analog under consideration here is of the order of $\alpha b$,
i.e. much larger, given that $b\gg 1$. We shall be interested in
$x_\perp=0$. Then, the $k_\perp$-integral in $\Delta A_0(x_3,0)$
becomes\bee\label{perturbation} \int_0^\infty \frac
{\exp\left(-\frac{k_\perp^2}{2m^2b}\right) \rmd k_\perp^2}
{(k_\perp^2+k_3^2)^2}=\frac{\exp
\left(\frac{k_3^2}{2m^2b}\right)}{2m^2b} {\rm
Ei}\left(-\frac{k_3^2}{2m^2b}\right)+\frac 1{k_3^2}. \eend Here Ei
is the exponential integral, and we have used Eq. 3.353.3 from
\cite{ryzhik}. When integrating this over $k_3$  we may pass to
the limit $(2m^2b/k_3^2)=(2eB/k_3^2)\rightarrow\infty$ in the
integrand, since the remaining integral \bee\label{uehling} \Delta
A_0(x_3,0)\simeq \frac{q\alpha b m^2}{4\pi^3}\int_{-\infty}^\infty
\rme^{-\rmi k_3x_3} T\left(\frac{k_3^2}{4m^2}\right)\frac{\rmd
k_3}{k_3^2}%=\frac{\alpha b}{4\pi^2}\int_0^\infty\frac{T(y)\rmd
%y}{y^{\frac 3{2}}}=\frac{\alpha b}{16}
 \eend converges both at
small and large integration variable (note the asymptotic behavior
of (\ref{T})). Next we use the integral representation (\ref{T})
and the residue method  to calculate the last integral. This leads
to the additive vacuum polarization correction to the Coulomb
potential in the form\bee\label{uehling2}\Delta A_0(x_3,0)\simeq
\frac{q\alpha b m}{8\pi^2}\int_0^{\pi/2}
\rme^\frac{-2m|x_3|}{\cos\phi}\cos^2\phi~\rmd\phi\eend Setting
$x_3=0$ in it we obtain for (\ref{constant}) \bee\label{C} C\simeq
\frac{\alpha b}{16}.
 \eend So, in the interval of magnetic fields indicated the constant $C$ in
the Laurent expansion (\ref{expand}) grows linearly with the
field, in contrast to the square root growth (\ref{C0})
characteristic of larger fields, as we saw in the previous
subsection. The correction (\ref{uehling2}) was used in
\cite{PRL07} to find the energy correction to (\ref{LE}) for $1\ll
b\ll 1000$.
\section*{Appendix II}

In this Appendix we solve, for asymptotically large magnetic fields
$b\gg (2\pi/\alpha)\sim 10^3,$ the eigenvalue problem inferred by
the \sch (\ref{sch}) with only the short-range part (\ref{short}) of
the modified Coulomb potential taken for $A_0(x_3,x_\perp=0)$. The
latter  is approximated, in accord with (\ref{expand}) with the
quadratic terms omitted, $C_{\perp}=C_{\parallel}=0$, as $eA_{\rm
s.r.}(x_3,0)\cong
-V(x_3)$\bee\label{broken}\hspace{-1cm}V(x_3)=\left\{\begin{tabular}{cc}
$-Z\alpha\left(\frac1{|x_3|}-2mC\right)$ &~~ for $\quad L_{\rm
B}<|x_3|<\overline{x_3}=\frac 1{2mC},$\\0 &~~ for $\qquad
|x_3|>\overline{x_3}=\frac 1{2mC}$\,,\end{tabular}\right.\eend where
the external-field-dependent constant $C$ (\ref{constant}) is given
by eq. (\ref{C0}). In the same way as in Sec. IIIB we may derive
that the potential (\ref{broken}) becomes the $\delta$-function in
the $b=\infty$ limit, with the coefficient, however, different from
the one in (\ref{deltax3}): \bee\label{brokdelta}
V(x_3)\left.\right|_{b=\infty}=-\frac{qe}{2\pi}\left[\ln
{\sqrt{b}\over 2C}-1 +{2C\over
\sqrt{b}}\right]\delta(x_3)\nonumber\hspace{7cm}\\=-\frac{qe}{2\pi}\left[\ln
\left({\pi\over 2\alpha}\right)^{1/2}-1+\left({2\alpha\over
\pi}\right)^{1/2}
\right]\delta(x_3)=-1.79\frac{qe}{2\pi}\delta(x_3).\hspace{3cm}\eend
The difference in coefficients is owing to the fact that we kept
only two terms in the expansion (\ref{expand}). In equality
(\ref{brokdelta}) Eq. (\ref{C0}) was used. The square root
asymptotic  dependence (\ref{C0}) of $C$ on the magnetic field is
crucial for the formation of the $\delta$-function limit of the
potential.

The approximation (\ref{broken}) replaces the curves in Fig.
\ref{fig:3} by continuous broken lines. The lowest energy state of
the \sch  ~(\ref{sch}) is determined by imposing the boundary
condition \cite{haines}
\bee\label{bound}\left.\frac{\rmd\Psi(x_3)}{\rmd
x_3}\right|_{x_3=L_{\rm B}}=0.\eend For the approximation
(\ref{broken}) to be meaningful it is necessary
that\bee\label{inequality} \overline{x_3}\gg L_{\rm B}.\eend With
Eq. (\ref{C0}) for $C$, this condition  reduces to the evident
inequality $137\pi/2\gg (0.9595)^2$ and is thus guaranteed.

Introducing the so-called quantum defect $\delta$ instead of the
eigen-energy $E$ according to the relation [remind that $a_{\rm
B}=(m\alpha )^{-1}$ is the Bohr radius] \bee\label{defect}
E-Z\alpha 2mC=-\frac{1}{2m\delta^2a_{\rm B}^2},\eend  and the new
variable $z=2x_3/\delta a_{\rm B}$ we obtain for (\ref{sch}) two
equations\bee\label{sch1}\hspace{-1.5cm} \frac{\rmd^2\Psi(z)}{\rmd
z^2}+\frac{Z\delta}{z}\Psi(z)-\frac 1{4}\Psi(z)=0,\qquad {\rm
for}\quad \frac{2L_{\rm B}}{\delta a_{\rm B}}\leq z\leq
\overline{z}=\frac 1{mC\delta a_{\rm B}}=\frac\alpha{C\delta}\eend
and\bee\label{sch2}\hspace{-1.5cm} \frac{\rmd^2\Psi(z)}{\rmd
z^2}+\frac{Z\delta^2C}{\alpha}\Psi(z)-\frac 1{4}\Psi(z)=0,\qquad
{\rm for}\quad  z\geq \overline{z}=\frac 1{mC\delta a_{\rm
B}}=\frac\alpha{C\delta}.\eend One should consider the couple of
Eqs. (\ref{sch1}) and (\ref{sch2}) with the boundary condition
\bee\label{bound2}\left.\frac{\rmd\Psi(z)}{\rmd
z}\right|_{z=\frac{2L_{\rm B}}{\delta a_{\rm B}}}=0\eend that
follows from (\ref{bound}), as an eigenvalue problem for
determining the quantum defect $\delta$ and hence the energy
(\ref{defect}). The general solution to the confluent
hypergeometric differential equation (\ref{sch1}) is \cite{jahnke}
\bee\label{general}\Psi =AW_{Z\delta,\frac
1{2}}(z)+BM_{Z\delta,\frac 1{2}}(z),\qquad \frac{2L_{\rm
B}}{\delta a_{\rm B}}\leq z\leq\overline{z},\eend where
$W_{Z\delta,\frac 1{2}}(z)$ is the Whittaker function, decreasing
at $z\rightarrow\infty$, while the other, linear independent
solution, growing at $z\rightarrow\infty$, $M_{Z\delta,\frac
1{2}}$ is expressed in terms of the confluent hypergeometric
function $\Phi$ as\bee\label{M} M_{Z\delta,\frac
1{2}}(z)=\rme^{-\frac z {2}}z\Phi(1-Z\delta,2;z),\eend and $A$ and
$B$ are constants.

We shall seek for the solution of the eigenvalue problem
(\ref{sch1}), (\ref{sch2}), (\ref{bound2}) in the region [serving
the asymptotically large magnetic fields considered here]
\bee\label{delta} \delta\gg\frac {\alpha}{C},\eend so that
$\overline{z}\ll 1.$ Therefore, only the small-distance behavior
of the fundamental solutions to Eq. (\ref{sch1}) will be
important. Referring to the asymptotic behavior of the solutions
at small $z$ \cite{haines,jahnke}
\bee\label{whittaker}\hspace{-2cm}W_{Z\delta,\frac 1{2}}(z)\cong
\frac{\exp
(-\frac{z}2)}{\Gamma(-Z\delta)}\left(-\frac{1}{Z\delta}+z[\ln z
+\psi(1-Z\delta)-\psi(1)-\psi(2)]+\mathcal{O}(z^2)\ln
z\right),\nonumber\\\hspace{-2cm}M_{Z\delta,\frac 1{2}}\cong z+
\mathcal{O}(z^2)\eend that retains the terms $z^0$, $z$ and $z\ln
z$ (here the logarithmic derivative $\psi$ of the Euler
$\Gamma$-function $\Gamma$ appears), Eq. (\ref{general}) is
matched continuously in the point
  $z=\overline{z}$ with the decreasing solution of the
\sch ~(\ref{sch2})\bee\label{decrease} \Psi(z)=\rme^{-f_\delta
z},\qquad z\geq\overline{z},\eend
 where $f_\delta =\sqrt{\frac
1{4}-\frac{\delta^2CZ}\alpha},$ and its first derivative over
$z$\bee\label{matching} \Psi(\overline{z})=1,\qquad
\left.\frac{\rmd\Psi(z)}{\rmd
z}\right|_{z=\overline{z}}=-f_\delta,\eend if the coefficients $A$
and $B$ in (\ref{general}) are taken as \bee\label{AB}A=\frac
1{W_{Z\delta,\frac 1{2}}(\overline{z})}
%(1-Z\delta\overline{z}\ln\overline{z})^{-1}
,\qquad %B=-\frac1{2}-f_\delta+Z\delta\ln\overline{z}.
B=-f_\delta-\frac 1{W_{Z\delta,\frac
1{2}}(\overline{z})}\left.\frac{\rmd W_{Z\delta,\frac
1{2}}(z)}{\rmd z}\right|_{z=\overline{z}}.\eend Keeping the
leading terms as $z\rightarrow 0$
($Z\delta\overline{z}\ln\overline{z}$ is neglected as compared to
1) we get from (\ref{whittaker})\bee\label{W}W_{Z\delta,\frac
1{2}}(\overline{z})=\frac 1{\Gamma(1-Z\delta)},\nonumber\\
\left.\frac{\rmd W_{Z\delta,\frac 1{2}}(z)}{\rmd
z}\right|_{z\rightarrow 0}=-\frac 1{2\Gamma(1-Z\delta)}+\frac{\ln
z +\gamma+\psi(1-Z\delta)-\psi(1)}{\Gamma(-Z\delta)},\eend where
$\gamma=-\psi(1)$ is the Euler constant. With these values, the
boundary condition (\ref{bound2}) results in the following
$algebraic$ equation for the quantum defect $\delta$
\bee\label{alg}\frac {f_\delta}
{Z\delta}=\ln\overline{z}-\ln\frac{2L_{\rm B}}{\delta a_{\rm
B}}\equiv -\ln (2mL_{\rm B}C).\eend
%If the term $\ln\overline{z}$ is omitted from equation (\ref{alg})
%and $\ln\delta$ is neglected as compared to $f_\delta/Z\delta$,
%its solution
%B}}{a_{\rm B}}\right)+\frac{4ZC}\alpha\right)^{-1},\eend
%after\bee\label{delta0}\delta^2=\left(4Z^2\ln^2\left(\frac{2L_{\rm
%being substituted into (\ref{defect}), leads
%and the value 1/2 is taken for $f_\delta$, the known equation
%\bee\label{known} \frac 1{2Z\delta}=-\ln\frac{2L_{\rm B}}{\delta
%a_{\rm B}}\eend leading
%to the known ground-state energy value without radiative
%corrections \bee\label{known2} E=E_0\equiv -\frac {2Z^2}{ma_{\rm
%B}^2}\ln^2\left(\frac{2L_{\rm B}}{a_{\rm B}}\right)\equiv  -
%2Z^2\alpha^2m\ln^2\left(\frac{\sqrt{b}}{2\alpha}\right), \eend %for the
%%case where the radiative corrections to the Coulomb potential are
%%disregarded is reproduced result
%from which the contribution of $C$ has canceled.
%For large magnetic fields under consideration here, the
%field-depending constant $C$ is sufficiently large, so that
%$\overline{z}=\alpha/C\delta$ is small and $\ln\overline{z}$
%cannot be disregarded as compared to $\ln\frac{2L_{\rm B}}{\delta
%a_{\rm B}}.$ Then, s
Solution to equation (\ref{alg}) is %obtained
%from (\ref{delta0}) by the replacement $\ln\left(\frac{2L_{\rm
%B}}{a_{\rm B}}\right)\rightarrow \ln\left(\frac{2L_{\rm
%B}}{\overline{z}a_{\rm B}}\right)$
\bee\label{delta1}\delta^2=\left(4Z^2\ln^2\left(2mCL_{\rm
B}\right)+\frac{4ZC}\alpha\right)^{-1},\eend and the ground-state
energy (\ref{defect}) is \bee\label{energy} E= -\frac {2Z^2}{ma_{\rm
B}^2}\ln^2\left(2mCL_{\rm B}\right)= -
2Z^2\alpha^2m\ln^2\left(\frac{\sqrt{b}}{2C}\right). \eend
%Note,that unlike the no-radiative-corrections case above, now we did
%not neglect $\ln\delta$.

It remains to make sure that the assumption (\ref{delta})
$\overline{z}=\alpha/C\delta\ll 1$ necessary for the use of the
asymptotic form of the solutions (\ref{whittaker}) made while
deriving equation (\ref{alg}) is indeed obeyed, once the quantum
defect $\delta$ is given by (\ref{delta1}). In other words, we
must check the strong
inequality\bee\label{check}2\left[\ln^2\left(\frac{2C}{\sqrt{b}}
\right)+\frac{C}{Z\alpha}\right]^{1/2}\ll \frac C{Z\alpha}.\eend
By solving the quadratic equation with respect to $C/Z\alpha$ it
becomes\bee\label{check2}1+\left[\ln^2\left(\frac{2C}{\sqrt{b}}\right)+1\right]^{1/2}\ll
\frac C{2Z\alpha}.\eend

For the fields so large that Eq. (\ref{C0}) holds for $C$, the
ln$^2$-term becomes independent of the magnetic field
\bee\label{ln}\ln^2\left(\frac{2C}{\sqrt{b}}\right)=\ln^2\sqrt{\frac\pi{2\alpha}}\equiv
7.213,\eend and the inequality (\ref{check2}) gives \bee\label{>>}
 b\gg \frac{8\pi\alpha
Z^2}{0.9}\left(1+\sqrt{1+\ln^2\sqrt{\frac\pi{2\alpha}}}\right)^2=8\pi\alpha
Z^2 16.6=3 Z^2.\eend This condition on the values of the magnetic
field is less restrictive than the condition of validity of Eq.
(\ref{C0}). Thus the inequality (\ref{delta}) is {\em aposteriory}
verified. Note, that
 also the inequality $Z\delta\ll 1$ %used for simplifying eqs.
%(\ref{asymp}), (\ref{asymp'}) is definitely
is satisfied  for the same fields, justifying the disregard of
$Z\delta\overline{z}\ln\overline{z}$ made when writing (\ref{W}). We
conclude that for asymptotically strong magnetic fields the
derivation that has led to Eq. (\ref{energy}) is justified, and the
ground-state energy acquires the magnetic-field-independent limiting
value\bee\label{energy2} E=-
2Z^2\alpha^2m\ln^2\sqrt{\frac\pi{2\alpha}}=-7.686\times
10^{-4}mZ^2=-389.3~{\rm eV}\times Z^2.\eend  %corresponding to the
%scaling regime discussed in Subsection 3.2.

\textit{Note added in proof.} Most recently a work \cite{tehran}
    appeared, where the modified Coulomb potential in a strong magnetic
    field calculated in \cite{arxive, PRL07} and in the present paper is
    also considered. In particular, in that work the long-range
    asymptotic behavior $|\bf{x}|^{-1}$ given as Eq. (\ref{llargex}) is
    supplemented by two next-to-leading terms of the order of
    $|\bf{x}|^{-3}$ and $|\bf{x}|^{-5}.$ Besides, the authors of
    \cite{tehran} found a small ($\sim \alpha/\pi$) anisotropic amendment
    to the simplified, Yukawa-like, form (\ref{yuk}) of the exact
    scaling equation (\ref{universal}).


\begin{thebibliography}{20}
\bibitem{to avoid} To avoid misunderstanding henceforce, note that
the singular behavior $1/r$ in the origin we refer to corresponds
to taking the limit $r=0$ while keeping $b=const$, whereas the
$\delta$-function behavior is explained by considering the limit
$b=\infty$ at constant $r$.
\bibitem{footnote1} In every figure throughout we plot the energy of an electron in the
field of a point charge instead of the potential, produced by this
charge, for convenience of considering the atomic spectra in the
last section. Note the corresponding change of sign of the curves.
\end{thebibliography}
\end{document}